\documentclass[journal]{IEEEtran}
\usepackage{graphicx}
\usepackage{tipa}
\usepackage{bbm}
\usepackage{mathrsfs}
\usepackage{pifont}
\usepackage{amsfonts}
\usepackage{cite,url,subfigure,epsfig,graphicx}
\usepackage{amssymb,amsmath}
\usepackage{indentfirst}
\usepackage{algorithmic}
\usepackage{algorithm}
\usepackage{pdfpages}
\usepackage{cite,url,subfigure,epsfig,graphicx}
\usepackage{times,verbatim,amsfonts,amsmath,color}
\usepackage{multirow}
\usepackage{booktabs}
\usepackage{adjustbox}

\IEEEoverridecommandlockouts
\makeatletter
\def\ps@headings{\def\@oddhead{\mbox{}\scriptsize\rightmark \hfil \thepage}\def\@evenhead{\scriptsize\thepage \hfil \leftmark\mbox{}}\def\@oddfoot{}\def\@evenfoot{}}
\makeatother 

\newtheorem{lemma}{ \textbf{Lemma}}

\newtheorem{theorem}{ \textbf{Theorem}}

\newtheorem{remark}{Remark}

\begin{document}

\title{A Platform-Free Proof of Federated Learning Consensus Mechanism for Sustainable Blockchains}

\author{
\IEEEauthorblockN{Yuntao~Wang\IEEEauthorrefmark{2}, Haixia Peng\IEEEauthorrefmark{3}, Zhou~Su\IEEEauthorrefmark{2}\IEEEauthorrefmark{1}, Tom H Luan\IEEEauthorrefmark{2}, Abderrahim~Benslimane\IEEEauthorrefmark{4}, and Yuan Wu\IEEEauthorrefmark{5}}\\ 
\IEEEauthorblockA{
\IEEEauthorrefmark{2}School of Cyber Science and Engineering, Xi'an Jiaotong University, Xi'an, China\\
\IEEEauthorrefmark{3}School of Information and Communications Engineering, Xi'an Jiaotong University, Xi'an, China\\
\IEEEauthorrefmark{4}Laboratory of Computer Sciences, Avignon University, France\\
\IEEEauthorrefmark{5}State Key Laboratory of Internet of Things for Smart City, University of Macau, Macau, China\\
\IEEEauthorrefmark{1}Corresponding author: Zhou~Su (zhousu@ieee.org)
}
\thanks{This work was supported in part by NSFC (nos. U20A20175, U1808207), the Fundamental Research Funds for the Central Universities, FDCT 0162/2019/A3 and SKL-IOTSC(UM)-2021-2023. This article has been accepted for publication in IEEE Journal on Selected Areas in Communications with DOI 10.1109/JSAC.2022.3213347}
}

\maketitle

\begin{abstract}
Proof of work (PoW), as the representative consensus protocol for blockchain, consumes enormous amounts of computation and energy to determine bookkeeping rights among miners but does not achieve any practical purposes. To address the drawback of PoW, we propose a novel energy-recycling consensus mechanism named platform-free proof of federated learning (PF-PoFL), which leverages the computing power originally wasted in solving hard but meaningless PoW puzzles to conduct practical federated learning (FL) tasks.
Nevertheless, potential security threats and efficiency concerns may occur due to the untrusted environment and miners' self-interested features. In this paper, by devising a novel block structure, new transaction types, and credit-based incentives, PF-PoFL allows efficient artificial intelligence (AI) task outsourcing, federated mining, model evaluation, and reward distribution in a fully decentralized manner, while resisting spoofing and Sybil attacks. Besides, PF-PoFL equips with a user-level differential privacy mechanism for miners to prevent implicit privacy leakage in training FL models. Furthermore, by considering dynamic miner characteristics (e.g., training samples, non-IID degree, and network delay) under diverse FL tasks, a federation formation game-based mechanism is presented to distributively form the optimized disjoint miner partition structure with Nash-stable convergence. Extensive simulations validate the efficiency and effectiveness of PF-PoFL.
\end{abstract}

\begin{IEEEkeywords}
Blockchain, AI-inspired consensus, federated learning, dynamic pool formation.
\end{IEEEkeywords}

\IEEEpeerreviewmaketitle

\section{Introduction}
Blockchain, as a disruptive next paradigm innovation, offers a decentralized solution to immutably store information, transparently execute transactions, and automatically establish trust in open and trustless environments \cite{9631953,8871181,9293091,9159929}. 
In blockchain systems, the consensus protocol is the core component which determines the system scalability, security, and consistency. The goal of consensus protocols is to ensure that all involved entities agree on an identical and consistent ledger without the need for a central authority.
The representative consensus protocol is the proof of work (PoW) \cite{8835227}, which is widely adopted in various blockchains such as Bitcoin and Ethereum. 
In PoW, miners compete for the bookkeeping rights and rewards by solving a cryptographic puzzle (which is hard to solve but easy to validate) through brute-forcing, i.e., \emph{mining} \cite{9165548}. Unfortunately, the mining process is computation-hungry and consumes an enormous amount of computation and energy in solving meaningless PoW puzzles, controversial to the trend of sustainable and environment-friendly technology. As reported, the annual electricity consumption of Bitcoin is comparable to that of Thailand, and the electricity that a single Bitcoin transaction consumes can power a U.S. household for about 3 weeks \cite{Digiconomist}.

To mitigate the energy waste in PoW, research efforts have been made from two aspects: \emph{energy saving} and \emph{energy recycling}.
Proof of stake \cite{9312484} and proof of space \cite{PoSpace} are two typical energy-saving alternatives, which conserve energy by reducing mining difficulty for wealthy stakeholders who invest more in cryptocurrency or storage space. An explicit side effect of these methods is the non-democracy and corresponding Matthew's effects as rich participants have a higher chance for bookkeeping to earn more rewards. Meanwhile, there still exists ``useless'' resource waste on computation or storage in reaching consensus, remaining far from being really ``useful''.

Another line of work studies the energy-recycling consensus approaches, where the wasted energy in solving hard but meaningless puzzles in PoW is recycled to perform practical useful works, known as proof of useful work (PoUW) \cite{Ball2017ProofsOU}.
In PoUW, PoW mining tasks can be replaced with practical computational problems such as searching prime chains \cite{Primecoin}, image segmentation \cite{9025613}, and all-pairs shortest path \cite{Ball2017ProofsOU}, which offers a suitable substitution of PoW for better sustainability of the blockchain system.
Recently, various works have attempted to design PoUW mechanisms using machine learning as the basis for useful tasks \cite{Lan2021PoLe,Chenli2019Energy}.
Essentially, similar to PoW mining, training artificial intelligence (AI) models usually costs considerable computing power and time while its verification process is much easier.
Besides, the current trend of going deeper in AI models inevitably results in the ever-increasing demand for computing power. Thereby, it is natural to reinvest the computation power in PoW to train valuable AI models and employ trained models as proofs for completing the consensus.

\begin{figure}[!tbp]
\centering 
  \includegraphics[width=9cm]{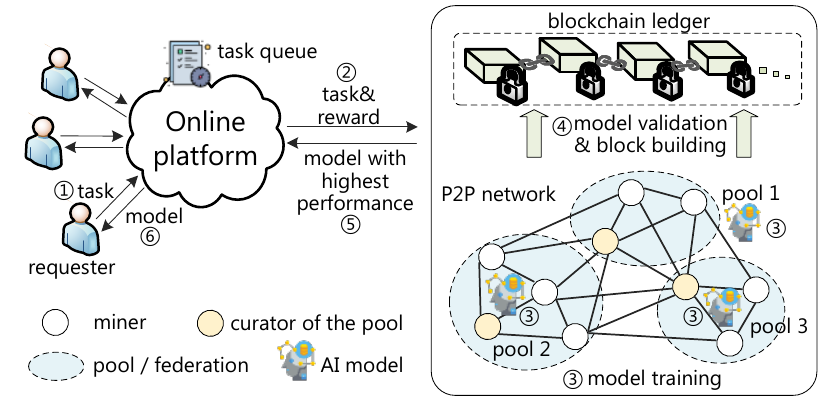}\\ 
  \caption{Architecture of the PoFL scheme \cite{9347812}. (\ding{172}: Requesters upload FL tasks with rewarding information to the FL platform; \ding{173}: The FL platform maintains a task queue and selects an unfinished task to start PoUW; \ding{174}: All members in a mining pool train the requested FL model and each pool uses the trained model as a proof; \ding{175}: The blockchain performs model validation and ranking; \ding{176}: The model with the highest performance is delivered to the platform and the pool who produces it earns the mining reward.)}\label{fig:introduction1}\vspace{-0.1cm}
\end{figure}

However, training a qualified AI model requires considerable training data as ``fuel", and a single miner or the task initiator may lack sufficient data samples on its device. Moreover, privacy concerns make it risky or even illegal in centralizing data samples from data owners (i.e., miners) for model training \cite{9293091}. As an attempt to fully unleash the power of AI using distributed data, Qu \emph{et al}. \cite{9347812} recently introduced a general proof of federated learning (PoFL) consensus framework by reinvesting miners' computing power in solving meaningless PoW puzzles to train high-quality federated learning (FL) models. FL is a distributed AI paradigm which usually contains (i) a crowd of data owners (e.g., \emph{miners}) who cooperatively train an AI model by sharing intermediate gradients and model parameters instead of their local datasets, and (ii) a central \emph{curator} for learning coordination. Under FL, miners' training datasets are kept on local devices and only the locally computed model updates are shared, thereby greatly alleviating data privacy concerns. As shown in Fig.~\ref{fig:introduction1}, in PoFL \cite{9347812}, users first outsource their FL tasks to the online third-party platform (e.g., Kaggle and Codalab), then all the members within a mining pool collaboratively train the requested FL model before the deadline, and finally different pools compete to earn rewards by using the trained models as proofs.

Although the work \cite{9347812} has made significant progress, there are still significant challenges that remain to be tackled.
First, the working of PoFL in \cite{9347812} heavily depends on the trusted third-party platform which hosts FL tasks issued by requesters, determines the order of tasks for mining, and delivers rewards to the winning pool in the race. The malfunction or breakdown of the platform can cause a system failure, which is essentially a single point of failure (SPoF). Besides, the platform may collude with certain mining pools and disclose the FL tasks to them in an early stage, so that they can start training before others and gain benefits. Therefore, \emph{it remains a concern to design a robust and fully decentralized PoFL scheme which gets rid of the central platform}. 
Next, the work \cite{9347812} directly leverages off-the-shelf pooled-mining structures in PoW-based blockchains to train FL models. Different from the homogeneous PoW tasks, miners usually have distinct training samples and non-IID degrees and correspondingly distinct contributions when conducting different FL tasks under PoFL, causing a more complex pool formation problem. \emph{There exists a challenge in dynamically forming optimized pooled structures with stable miner-pool association for diverse FL tasks}.
Furthermore, in the fully distributed and autonomous blockchain system, well-designed incentives for self-interested miners are benign impetuses to build a healthy consensus system with enhanced efficiency. Due to the intrinsic selfishness and behavior diversity of miners, \emph{how to offer precise incentives to heterogenous miners for optimized consensus process is a challenging issue}.

\begin{figure}[!t]
\centering 
  \includegraphics[width=9cm]{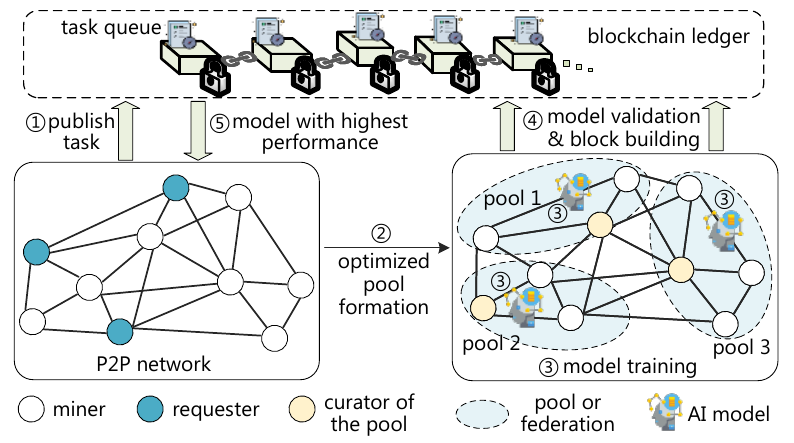}\\ 
  \caption{Architecture of our PF-PoFL scheme. (\ding{172}: Requesters upload FL tasks with rewards to the blockchain which maintains an unfinished task queue; \ding{173}: Miners dynamically form the optimized pool structure to perform PoUW for an unfinished task; \ding{174}: All members in a pool train the requested FL model in a differentially private manner and each pool uses the trained model as a proof; \ding{175}: Selected validators in blockchain perform model validation and ranking for the uncompleted task; \ding{176}: The model with the highest performance is delivered to its requester and the pool who produces it earns the mining reward.)}\label{fig:introduction2}\vspace{-0.1cm}
\end{figure}

To address these issues, we propose a novel platform-free proof of federated learning (PF-PoFL) scheme to build a sustainable and robust blockchain ecosystem by removing the central platform and forming a dynamically optimized pooled structure for AI model training.
As shown in Fig.~\ref{fig:introduction2}, PF-PoFL involves three types of blockchain nodes: (i) \emph{requesters} that produce FL tasks, (ii) model \emph{trainers} (consisting of \emph{curators} and \emph{miners}) that train FL models within a pool, and (iii) \emph{validators} (selected from miners) that are in charge of model rankings and proposing new blocks.
Specifically, requesters directly announce their FL tasks with rewards to the blockchain, and a group of validators collectively maintain an ordered queue of unfinished tasks.
For an unfinished task, a crowd of miners with distinct training data size, non-IID degree, and network delay dynamically form a stable pooled structure via the federation formation game.
Next, coordinated by the corresponding curator, different pools compete to train the FL model and the pool with the best model earns the task reward via the consensus process. The validators are evenly compensated with the transaction fees.
The main contributions of this work are summarized below.

\begin{itemize}
  \item \emph{Sustainable and robust blockchain framework}. We propose PF-PoFL, an energy-recycling consensus framework which leverages the computing power of blockchain for efficient AI task outsourcing and reward distribution in a fully decentralized manner. Under PF-PoFL, we devise a novel block structure, new transaction types, and credit-based incentives to support FL model ranking and realize efficient consensus in the blockchain.
  \item \emph{User-level privacy-preserving federated mining}. We enhance the record-level differential privacy (DP) by designing a user-lever DP (UDP) algorithm to prevent the leakage of a user's whole data records during federated mining. Then, we rigorously analyze the sensitivity bound of a global aggregation and prove that the federated mining process satisfies $(\epsilon,\delta)$-UDP by adapting the variance of Gaussian noise added by curators.
  \item \emph{Game-theoretical optimized pool formation}. We formulate the disjoint pool formation problem for performing distinct FL tasks as a social welfare maximization problem, and propose a distributed federation formation game-based algorithm, which converges to Nash-stable equilibrium. By introducing the switch gain, both miners and pools execute an iterative two-sided matching process, where each miner decides its switch strategy to switch to the optimal pool while each pool decides its admission strategy to accept the optimal miner.
  \item \emph{Extensive simulations for performance evaluation}. We evaluate the efficiency and effectiveness of PF-PoFL using extensive simulations. Numerical results show that the PF-PoFL can attain not only stable blockchain throughput and desirable model performance with UDP guarantees, but also lower computation cost in reaching consensus and higher efficiency in federated mining, compared with conventional approaches.
\end{itemize}

The remainder of the paper is organized as follows. In Section \ref{sec:RELATEDWORK}, we review the related works. Section \ref{sec:SYSTEMMODEL} introduces the system model. Section \ref{sec:FRAMEWORK} provides a detailed construction of the PF-PoFL framework.
We analyze the game-theoretical stable pool formulation process in Section \ref{sec:COALITION}.
We evaluate the performance of our proposed framework in Section \ref{sec:SIMULATION}. Section \ref{sec:CONSLUSION} concludes this paper. 

\section{Related Works}\label{sec:RELATEDWORK}

In the literature, there has been an increasing interest in designing energy-recycling consensus mechanisms by substituting the hash puzzles in PoW with valuable works which are hard to solve but easy to verify.
Initially, early efforts focus on replacing the PoW nonce with practical computational problems.
Ball \emph{et al}. \cite{Ball2017ProofsOU} proposed a novel concept proof of useful work (PoUW) to replace the PoW mining with mathematic problems such as all-pairs shortest path and orthogonal vectors.
King \cite{Primecoin} proposed a novel cryptocurrency named Primecoin to generate scientific value in the mining process, where miners are required to search long prime chains as proofs of work with difficulty adjustability and non-reusability.
Shoker \cite{8171383} devised a sustainable consensus protocol named proof of exercise, in which miners choose to solve matrix-oriented computation issues hosted on a third-party platform instead of calculating hashes in PoW.
Daian \emph{et al}. \cite{Daian2017SPP} developed PieceWork to mitigate the energy waste in PoW by solving outsourced tasks such as denial-of-service (DoS) defense and spam prevention.

Recently, many attempts have been made in the integration of AI and blockchain by exploiting the computing capacity of blockchain for AI model training.
Li \emph{et al}. \cite{9025613} proposed a PoUW mechanism to reuse the wasted computing power in PoW for deep learning (DL)-based biomedical image segmentation to help clinical diagnosis.
Chenli \emph{et al}. \cite{Chenli2019Energy} developed a proof of deep learning (PoDL) consensus protocol by forcing miners to perform DL training and employing the trained models as proofs to earn rewards for outsourced DL tasks.
Lan \emph{et al}. \cite{Lan2021PoLe} devised an efficient proof of learning consensus mechanism to ensure data integrity and prevent malicious behaviors without sacrificing model performance.
Nevertheless, a single user may lack sufficient training data, and the centrally aggregated training/test dataset for public model training/verification may sacrifice user privacy and frustrate the adoption of blockchain.

Distinguished from the above works, our work combines blockchain and AI under the FL paradigm in a fully decentralized manner with privacy preservation and dynamic pool formation functions.

\section{System Model}\label{sec:SYSTEMMODEL}
We introduce the system model in this section, which consists of the blockchain model and adversary model. Then, we discuss the desired properties of PF-PoFL. 

\subsection{Blockchain Model}\label{subsec:networkmodel}
\begin{figure}[!t]
\centering\setlength{\abovecaptionskip}{-0.cm}
  \includegraphics[width=9cm]{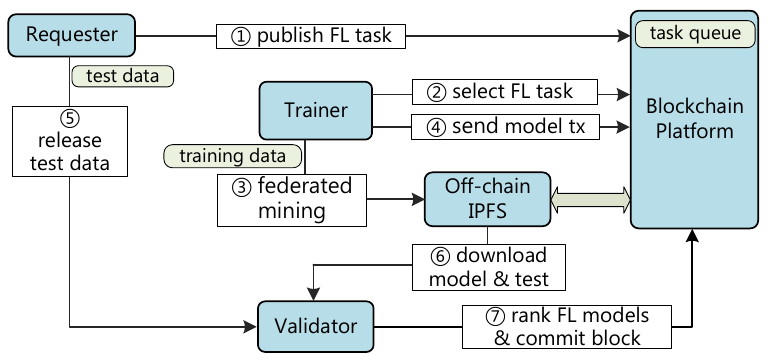}\\
  \caption{Entities and workflow of PF-PoFL.}\label{fig:model}\vspace{-0.1cm}
\end{figure}

The proposed decentralized blockchain network in this paper mainly consists of three types of nodes: requesters, trainers, and validators, as illustrated in Fig.~\ref{fig:model}.
\begin{itemize}
  \item \emph{Requesters}. Every node in the blockchain can act as a task requester to publish its FL tasks (e.g., semantic analysis and biomedical image recognition) to the blockchain platform with specific completion deadline and performance requirements, as well as the corresponding rewards as incentives. Let $\mathcal{R} = \{1,\cdots,r,\cdots,R\}$ be the set of task requesters in the blockchain network. The set of published tasks and unfinished tasks in the blockchain are denoted as $\mathcal{T} = \{1,\cdots,t,\cdots,T\}$ and $\mathcal{T}_u = \{1,\cdots,\tau,\cdots,T_u\}$, respectively.
  \item \emph{Trainers}. Analogy to the pooled-mining in PoW-based blockchains such as Bitcoin, a set of model trainers in PF-PoFL can form the pooled (or federated) structure \cite{9347812}, which is featured with intra-pool cooperation and inter-pool competition. A set of pools or federations are formed in the blockchain network, and the set of which is defined as $\Im = \{\Im_1,\cdots,\Im_j,\cdots,\Im_J\}$. In each pool/federation $\Im_j\in \Im$ managed by the corresponding curator $\phi_j$, a group of miners, denoted as $\mathcal{M}_j = \{1,\cdots,m,\cdots,M_j\}$, cooperatively train a qualified FL model for an unfinished task $\tau\in \mathcal{T}_u$ within the pool, and compete to earn the task reward with other pools in $\Im\backslash\{\Im_j\}$. 
  \item \emph{Validators}. In PF-PoFL, validators are a group of consensus nodes, denoted as $\mathcal{V} = \{1,\cdots,v,\cdots,V\}$, which are responsible for transaction verification and new block construction. An ordered queue of uncompleted tasks, denoted as $\widetilde{\mathcal{T}_u} = \mathrm{ordered}(\mathcal{T}_u)$, is collectively maintained by validators and is updated before the block height grows. Validators also perform the evaluation and ranking operations for released FL models, which are trained by each pool, via the consensus process.
\end{itemize}

In the blockchain network, a node can simultaneously perform all three roles or two of them. The set of all nodes in blockchain is denoted as $\mathcal{N} = \{1,\cdots,n,\cdots,N\}$.
The blockchain ledger $\mathcal{B}$ is composed of a series of growing hash-chained blocks, i.e., $\mathcal{B}=\{\mathcal{B}_{i} \leftarrow \mathcal{B}_{i-1}\leftarrow \cdots \leftarrow \mathcal{B}_{0}\}$, where $\mathcal{B}_{0}$ is called the genesis block, and $\mathcal{B}_{i}$ is the latest block with height $i$. Specifically, $\mathcal{B}_{i}$ contains two parts (i.e., block header and block body) and can be denoted by
\begin{align}\label{eq:block}
\mathcal{B}_{i}=\Big\{\underbrace{\mathrm{Hash}(\mathcal{B}_{i-1})||T_{\mathrm{stamp}}|| \mathrm{root}_i|| \mathrm{meta}_i}_{{\mathrm{header}}}||\underbrace{\mathcal{TX}_i}_{{\mathrm{body}}}\Big\}.
\end{align}
In Eq. (\ref{eq:block}), ${\mathcal{TX}_i}$ is a set of timestamped transactions in the block, which is organized into a Merkle tree structure with root value $\mathrm{root}_i$. $\mathrm{Hash}(\mathcal{B}_{i-1})$ is a cryptographic hash of the previous block to maintain a chained structure. $T_{\mathrm{stamp}}$ is the timestamp for block creation. $\mathrm{meta}_i$ is the metadata for validating and ranking the trained FL models, which ensures the validity of block $\mathcal{B}_{i}$ (details are shown in Sect.~\ref{subsec:scheme5}). The metadata $\mathrm{meta}_i$ can be analogical to the random nonce in Bitcoin, which serves a verifiable proof of the work.

\subsection{Adversary Model and Assumptions}\label{subsec:threatmodel}
In the system, model trainers (i.e., the curator $\phi_j$ and miners in $\mathcal{M}_j$) in each pool $\Im_j\in \Im$ are assumed to be \emph{semi-honest}. In other words, they will honestly obey the FL procedure in AI model training, while the miners may plagiarize others' trained FL models and the curator may be curious about miners' privacy information and refuse to pay at last.
Besides, the blockchain platform is assumed to be \emph{secure} and its stored transactions are \emph{immutable} and \emph{non-repudiable}.
Particularly, the following four kinds of attacks are considered:

\textbf{Spoofing Attack}. The model trainers may carry out spoofing attacks to gain high illegal profits. For example, they may publish a perfect model by training on the released testing data to win the competition with paltry cost. They could also plagiarize the FL models trained by other pools \cite{9347812}.

\textbf{Sybil Attack}. During FL model training, malicious trainers may launch Sybil attacks by submitting multiple meaningless local updates (to deteriorate model performance) or uploading the same local update multiple times (to earn more benefits). After FL model training, malicious pools may submit multiple meaningless models (to waste system resources) or send the same model multiple times (to earn more benefits), using Sybil identities. Besides, adversarial validators may create an arbitrary number of Sybil pseudonyms to influence the process of reaching an agreement on blockchain ledgers \cite{Gilad2017Algorand}.

\textbf{Untrusted Third-Party Platform}. In traditional PoFL \cite{9347812}, the central FL platform operated by third parties is responsible for hosting and ranking the FL tasks to be performed, as well as the payment delivery, whose malfunction or breakdown can result in a SPoF. The malicious platform may also collude with part of miners to conduct market manipulation by revealing FL tasks to them in advance. In addition, the platform may be compromised by attacks such as DDoS, causing leakage of miners' sensitive information that is stored in it.

\textbf{Implicit Privacy Leakage}. According to \cite{Zhu2019NEURIPS,8737416,7958568}, user privacy (e.g., the participation status in a task and the data used for training) may still be leaked unconsciously under FL in sharing raw local updates through differential attacks and advanced inference attacks.

\subsection{Desired Properties}\label{subsec:designgoal}
The design goal of the PF-PoFL is to achieve the following desirable properties simultaneously.

\textbf{1) Fully decentralized operation}. The whole process of PF-PoFL including FL task outsourcing, AI model training, model evaluation and ranking, and reward distribution should be operated in a fully decentralized paradigm.

\textbf{2) Dynamically optimized pool structure}. PF-PoFL aims for an optimized pool structure to competitively train FL tasks by considering dynamic user characteristics such as miners' cooperation and competition.

\textbf{3) Spoofing and Sybil prevention}. PF-PoFL should resist the Sybil identities and defend against spoofing attacks.

\textbf{4) User-level differential privacy (UDP)}. Existing works in FL mainly focus on the record-level differential privacy (RDP) \cite{Abadi2016Deep,9567711,9069945}, namely, whether a certain data sample is used for training. Different from them, we focus on the user-level privacy protection, which offers stronger provisions to protect the whole data samples of a user, i.e., the learned model does not leak whether a user participates in FL or not. 

\section{PF-PoFL: Platform-Free Proof of Federated Learning Framework}\label{sec:FRAMEWORK}

\subsection{Design Overview}\label{subsec:scheme0}
The workflow of PF-PoFL consensus framework consists of four successive phases: (i) FL task publication, (ii) FL model training, (iii) model ranking and rewarding, and (iv) block building and commit.
\begin{itemize}
  \item \emph{FL task publication}. Requesters publish their FL tasks $\mathcal{T}$ with rewards to the blockchain platform which maintains an ordered queue of uncompleted tasks $\widetilde{\mathcal{T}_u}$.
  \item \emph{Federated mining with UDP}. A group of trainers form an optimized pool/federation structure $\Im$ and choose an uncompleted task $\tau\in \widetilde{\mathcal{T}_u}$ to perform model training within the pool $\Im_j \in \Im$ under FL in a privacy-preserving manner. Different pools compete to train and submit the best FL model before the task deadline.
  \item \emph{Model ranking and rewarding}. The validators mutually execute the model ranking contract to evaluate and rank the trained FL models. Besides, validators perform the block rewarding contract to automatically enforce financial rewarding and update credits for participants. 
  \item \emph{Block building and commit}. Each validator executes the credit-based Algorand Byzantine agreement protocol to efficiently reach consensus on the new block to be added into the blockchain with deterministic finality.
\end{itemize}

In PF-PoFL, three types of blockchain transactions are implemented as follows.
\begin{itemize}
  \item \textbf{Payment Transaction} refers to transferring or redeeming the cryptocurrency from one node to another node for task payment, participation fee, etc.
  \item \textbf{Task Publication Transaction} in which a task requester proposes its FL tasks for learning competition.
  \item \textbf{FL Model Transaction} in which a training pool uploads its trained FL model as a solution for a particular FL task.
\end{itemize}

\subsection{System Initialization}\label{subsec:scheme1}
For system initialization, a bilinear pairing $e: \mathbb{G}_1 \times \mathbb{G}_2 \rightarrow \mathbb{G}_3$ is chosen, where $\mathbb{G}_1$, $\mathbb{G}_2$, and $\mathbb{G}_3$ are groups of prime order $q$. The generators of $\mathbb{G}_1$ and $\mathbb{G}_2$ are $g_1$ and $g_2$, respectively. Two hash functions $\mathrm{H}_1:\{0,1\}^*\rightarrow \mathbb{G}_2$ and $\mathrm{H}_2:\{0,1\}^*\rightarrow \mathbb{Z}_q$ are selected as random oracles. Taking a security parameter $\varphi$ as an input, the bilinear group generator $\mathbb{G}(\varphi)$ outputs the system parameters as ${params} = (q, g_1, g_2,\mathbb{G}_1, \mathbb{G}_2, \mathbb{G}_3, e)$.
Both $\mathbb{G}(\cdot)$ and ${params}$ are recorded in the genesis block.

Based on these public information, each participant $n \in \mathcal{N}$ sets up its secret key in the blockchain via $sk_n \xleftarrow{\mathrm{R}} \mathbb{Z}_q$ and generates its public key via $pk_n \leftarrow g_2^{sk_n}$, where $\xleftarrow{\mathrm{R}}$ represents randomly sampling. The wallet address of each node $n$ can be computed as $wa_n \leftarrow \mathrm{H}_2(pk_n)$. Each blockchain node $n$ stores $(sk_n,pk_n,wa_n)$ into its local tamper-proof device.

\subsection{FL Task Publication}\label{subsec:scheme2}
In PF-PoFL, the machine learning problems such as sentiment analysis and image classification are produced by the requesters along with the corresponding rewards. Specifically, each requester $r \in \mathcal{R}$ can publish an AI task $t\in \mathcal{T}$ at any time by sending a task publication transaction $tx_t$ to the blockchain. The detailed transaction format is as below.
\begin{itemize}
  \item Inputs:\\
    --~Task reward $p_t$ and task hosting fee $\xi_1$ of task $t$;\\
    --~Initial AI model parameters $\Theta _{t}(0)$;\\
    --~Performance metrics $\psi$ to evaluate model performance such as accuracy;\\
    --~The testing release block height $i_{t}$, i.e., the deadline (measured by block height \cite{8783030}) to release test dataset $\mathcal{D}_{\mathrm{test}}$;\\
    --~Cryptographic hash of the test dataset $\mathrm{dID} \leftarrow  \mathrm{Hash}(\mathcal{D}_{\mathrm{test}})$.
  \item Outputs: Task publication transaction $tx_t$:
\end{itemize}
\begin{align}\label{eq:taskTX}\resizebox{1.0\hsize}{!}{$
tx_t\!=\!\Big\{tx_t^\mathrm{ID},p_t,\xi_1,\Theta _{t}(0), \psi,i_{t},\mathrm{dID},T_{\mathrm{stamp}},{pk_r},\mathrm{Sig}_{sk_r}(tx_t^\mathrm{ID})\Big\}\!,\!$}
\end{align}
where $tx_t^\mathrm{ID}$ is the unique transaction identifier, i.e., hash of the transaction $tx_t$. $T_{\mathrm{stamp}}$ is the timestamp for transaction generation. $\mathrm{Sig}_{sk_r}(tx_t^\mathrm{ID})\leftarrow \mathrm{H}_1(tx_t)^{sk_r}$ is the digital signature of requester $r$ on the transaction. A small task hosting fee $\xi_1$ is utilized to encourage the validators to package $tx_t$ into the block and prevent malicious requesters from sending multiple meaningless tasks to cause a DoS.

The task publication transaction $tx_t$ also contains a payment transaction in which the requester transfers the task reward $p_t$ and task hosting fee $\xi_1$ to an escrow address $esAdd$ under public supervision. The formal transaction format is as follows.
\begin{itemize}
  \item Inputs: Task identifier $tx_t^\mathrm{{ID}}$ and user identifier ${pk_r}$.
  \item Outputs: Payment transaction $tx_p$:
\end{itemize}
\begin{align}\label{eq:taskTX}
tx_p\!=\!\Big\{tx_p^\mathrm{{ID}},tx_t^\mathrm{{ID}},p_t,\xi_1,T_{\mathrm{stamp}},{pk_r},\mathrm{Sig}_{sk_r}(tx_p^\mathrm{{ID}})\Big\}\!,\!
\end{align}
where $tx_p^\mathrm{ID}$ is the identifier (i.e., hash) of the transaction $tx_p$.
Here, if the requester fails to release the test dataset for task $t$ after the testing release block height $i_{t}$, it will lose these cryptocurrencies.

After receiving the task and payment transactions, validators mutually validate each of them via the function ${verify}(tx_t)$ by checking: (i) the task requester $r$ holds enough funds to pay the task hosting fee $\xi_1$ and reward $p_t$; and (ii) the difference between the testing release block height $i_{t}$ and the current block height $i$ is positive and larger than the minimum model training period preset by the system. If verification passes, we have ${verify}(tx_t)=true$, otherwise ${verify}(tx_t)=false$.
The invalid transactions are discarded and only valid ones are forwarded to the network.

For all valid tasks to be completed, each validator $v\in \mathcal{V}$ independently ranks these tasks in ascending order of timestamp and maintains a ready queue $\widetilde{\mathcal{T}_u}$ of uncompleted tasks. 
Generally, the task reward reflects the importance and urgency of a machine learning task.
As nodes in blockchain are selfish and rational, it is reasonable to assume that the priority of each running task $t$ for miners is determined by joint effect of the announced task reward $p_{t}$ and the remaining task training duration $i_{t}-i$. In other words, miners prefer FL tasks with higher rewards and longer remaining training duration. For the case that multiple tasks have the same priority, these tasks are ordered by their publication timestamp, i.e., the earlier-arrived one gets a better ranking.
After reaching an agreement among validators, the ready queue $\widetilde{\mathcal{T}_u}$ is updated and recorded in the blockchain.
In practice, we further implement a soft handover mechanism in which the PF-PoFL will be rolled back to PoW once the ready queue is empty.

\subsection{Federated Mining with UDP}\label{subsec:scheme3}
\begin{figure}[!t]
\centering\setlength{\abovecaptionskip}{-0.1cm}
  \includegraphics[width=9cm]{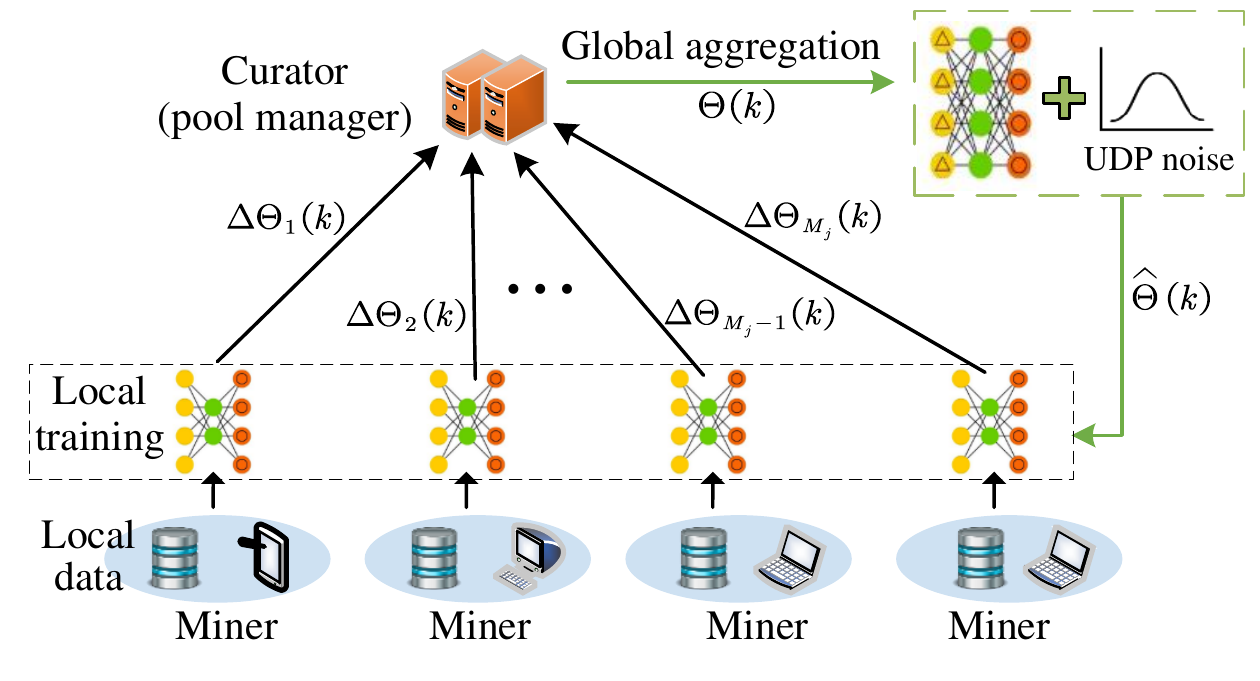}\\
  \caption{Illustration of federated mining with UDP in a pool under FL.}\label{fig:FLMining}\vspace{-0.1cm}
\end{figure}

Currently, there is a trend of pooled-mining in blockchain, which has a similar clustering structure with FL \cite{9347812}. In this work, we study PF-PoFL under the pooled-mining structure, as shown in Fig.~\ref{fig:FLMining}. The detailed federated mining process consists of the following steps.

\emph{\textbf{Step 1: }Task selection \& pool formation.}
Each trainer can pick any task $\tau \in \widetilde{\mathcal{T}_u}$ published on the blockchain which is still in its training phase (i.e., $i\le i_{\tau}$). Let $\mathcal{M}_\tau\subseteq \mathcal{M}$ be the set of trainers or miners that involve in training FL task $\tau$.
All trainers in the set $\mathcal{M}_\tau$ dynamically form an optimized stable pooled structure $\Im^*$ (details are shown in Sect.~\ref{sec:COALITION}) in a distributed manner.
For each pool $\Im_j\in \Im^*$, a group of trainers within the pool collaboratively train a qualified AI model for that task under the FL paradigm using their private data.
Besides, in each pool $\Im_j\in \Im^*$, a curator $\phi_j$ is randomly selected from the pool members whose experienced network latency is less than a predefined threshold $D_{\mathrm{th}}$.
Distinguished with the PoW protocol, in our PF-PoFL consensus protocol, different mining pools can concurrently train on different FL tasks that are in the training phase.

\emph{\textbf{Step 2: }FL model training with UDP within a pool.}
In pooled-mining under FL, as depicted in Fig.~\ref{fig:FLMining}, each pool member (i.e., miner) computes the intermediate gradients (i.e., local model update) of the current global model based on its local dataset and uploads them to the curator (i.e., pool manager) for periodic global aggregation (i.e., global model update).
Let $\Psi_{j,\tau}^{\mathrm{global}}$ be the number of communication rounds in training task $\tau$.
Initially, each miner in pool $\Im_j$ initializes its local model parameters as the initial model parameters $\Theta _{t}(0)$ downloaded from the transaction $tx_\tau$.
At $k$-th communication round ($k=1,2,\cdots,\Psi_{j,\tau}^{\mathrm{global}}$), it consists of a parallel local training stage on miners followed by a global aggregation and perturbation stage on the curator.

\emph{1) Local training at miner's side}. After receiving the previous global model $\widehat{\Theta}(k-1)$, each miner $m\in \Im_j$ individually calculates its local model $\Theta _{m}(k)$ using its local dataset via mini-batch stochastic gradient descent (SGD) with learning rate $\eta$, i.e., 
\begin{align}\label{eq:localtraining}
\Theta _{m}(k) \leftarrow \widehat{\Theta}(k-1) - \eta \nabla{\mathcal{L}}\big( \widehat{\Theta}(k\!-\!1) \big),
\end{align}
where $\nabla{\mathcal{L}}$ is the gradient of loss function ${\mathcal{L}}(.)$ on the mini-batch.
Then miner $m$ calculates the local updates by
\begin{align}\label{eq:localupdate}
\Delta\Theta _{m}(k) \leftarrow \Theta _{m}(k) - \widehat{\Theta}(k-1),
\end{align}
and uploads $\Delta\Theta _{m}(k)$ to the curator.

\emph{2) Global aggregation and perturbation at curator's side}. The curator $\phi_j$ aggregates the local updates into a global model $\Theta(k)$. Besides, to ensure user-level privacy preservation under FL, the curator applies a randomized function $\mathcal{F}$ to add the UDP noise to the sum of scaled local updates.

The above training process is repeated until a desirable accuracy of the global model is attained or the communication round reaches its maximum value. In the following, we first give the definition of UDP and then present the detailed UDP implementation mechanism.

\emph{\textbf{Definition 1}. (User-level Differential Privacy (UDP)):}
A randomized function $\mathcal{F}: \mathcal{D}\rightarrow \mathcal{R}$ satisfies $(\epsilon,\delta$)-UDP if for any two user-adjacent datasets $y, y' \in \mathcal{D}$ and for any subset of outputs $\mathcal{Y} \subseteq \mathcal{R}$, the following inequality holds:
\begin{align}\label{eq:UDP}
\Pr \left[ {\mathcal{F}\left( y \right) \in \mathcal{Y}} \right] \le e^{\epsilon }{\Pr \left[ {\mathcal{F}\left( {y'} \right) \in \mathcal{Y}} \right]} + \delta,
\end{align}
where $\mathcal{D}$ and $\mathcal{R}$ are the domain (e.g., possible training datasets) and range (e.g., possible trained global models) of $\mathcal{F}$, respectively. Two datasets $y, y'$ are user-neighboring if $y$ can be formed from $y'$ by removing or adding all data samples related to a single miner.

\begin{remark}
In conventional record-level DP (RDP) mechanisms \cite{Abadi2016Deep,9567711,9069945}, the datasets $y, y'$ are defined to be record-neighboring, i.e., $y$ can be formed from $y'$ by removing or adding a single data record. Instead of protecting a single data sample of the miner, we design a UDP mechanism $\mathcal{F}$ to protect the miner's whole data samples in the training process. As such, any adversary observing the published final model cannot deduce the participation of any specific miner and the usage of any specific data samples in model training with strong probability.
\end{remark}

The Gaussian mechanism is adopted to design such function $\mathcal{F}$ by sanitizing the sum of all updates with low utility decrease. To enforce the bounded sensitivity of local updates, the curator $\phi_j$ scales each local update by
\begin{align}\label{eq:scaledupdate}
\Delta\tilde{\Theta} _{m}(k) \leftarrow \frac{\Delta\Theta _{m}(k)}{\max\left\{1,\frac{{||\Delta\Theta _{m}(k)}||_2}{A}\right\}}.
\end{align}
Thereby, ${||\Delta\tilde{\Theta} _{m}(k)}||_2 \leq A$ can be ensured, and the sensitivity in the summing operation of all scaled updates is upper bounded by $A$. According to \cite{geyer2017clientDP}, we choose $A = \mathrm{median}\{{||\Delta\Theta _{m}(k)}||_2:m\in \mathcal{M}_j\cup\{\phi_j\}\}$.

The Gaussian noise scaled to sensitivity $A$ with zero mean, i.e., $\mathcal{G}(0,\sigma^2 A^2)$, is added to the sum of the scaled updates to prevent individual privacy leakage, i.e.,
\begin{align}\label{eq:globalagg}
\widehat{\Theta}(k) \!\leftarrow\! \widehat{\Theta}(k\!-\!1) \!+\! \frac{1}{M_j\!+\!1}\left[\sum_{m=1}^{|\mathcal{{M}}_j\cup\{\phi_j\}|}{\Delta\tilde{\Theta} _{m}(k)} \!+\! \mathcal{G}(0,\sigma^2 A^2) \right]\!,\!
\end{align}
where the parameter $\sigma$ controls the scale of Gaussian noise.
After that, the curator $\phi_j$ delivers the perturbed global model $\widehat{\Theta}(k)$ to all miners in the pool.

\emph{\textbf{Step 3: }FL model submission.}
Once pool $\Im_j$ accomplishes the FL learning process, the curator $\phi_j$ as the pool manager submits its trained final model $\Theta_{j,\tau}$ by sending a FL model transaction, whose formal format is as follows.
\begin{itemize}
  \item Inputs:\\
    --~Hash of the trained model $\Theta_{j,\tau}$, i.e., $\mathrm{mID}\leftarrow \mathrm{Hash}(\Theta_{j,\tau})$;\\
    --~Reference score $rs$ in model training phase;\\
    --~Participation fee $\xi_2$;\\
    --~Aggregated public key $apk_j$ of trainers in the pool;\\
    --~Multi-signature $\mathrm{MulSig}(tx_m^\mathrm{ID})$ of trainers in the pool.
  \item Outputs: FL model transaction $tx_m$:
\end{itemize}
\begin{align}\label{eq:taskTX}
tx_m\!=\!\Big\{tx_m^\mathrm{ID},tx_{\tau}^\mathrm{ID},\mathrm{mID},rs,\xi_2,T_{\mathrm{stamp}},{apk_j},\mathrm{MulSig}(tx_m^\mathrm{ID})\Big\}\!,\!
\end{align}
where $tx_m^\mathrm{ID}$ is the transaction identifier.
Notably, each mining pool only submits the hash value of the trained AI model (i.e., $\mathrm{mID}$) at this stage, to prevent from being plagiarized by other rivals. Once the test dataset is released, all involved pools will upload their trained models to an off-chain IPFS platform.
Besides, the training score $rs$ can be a reference value of model performance and is not considered for model ranking. In addition, a small participation fee $\xi_2$ is involved in $tx_m$ to defend against Sybil attacks conducted by malicious trainers/pools during/after FL model training.
The participation fee $\xi_2$ is evenly assigned to all members within each pool, which can be redeemed to the trainers whose model performance is above a certain threshold in consensus model ranking.
Similar to transaction $tx_\tau$, a payment transaction is involved, i.e.,
\begin{align}\label{eq:taskTX}
tx_p\!=\!\Big\{tx_p^\mathrm{{ID}},tx_m^\mathrm{ID},\xi_2,T_{\mathrm{stamp}},{pk_{\phi_j}},\mathrm{Sig}_{sk_{\phi_j}}(tx_p^\mathrm{{ID}})\Big\}.
\end{align}

For trainers in pool $\Im_j$, their aggregated public key can be derived as
\begin{align}
apk_j = \prod\limits_{n \in \mathcal{M}_j \cup \{\phi_j\}} {pk_n^{ \mathrm{H}_2(pk_n,\{pk_1,\cdots,pk_{M_j},pk_{\phi_j}\})}}.
\end{align}
We define $\varsigma_{n}\leftarrow { \mathrm{H}_2(pk_n,\{pk_1,\cdots,pk_{M_j},pk_{\phi_j}\})}$ to ease the formulation. For each pool member $n\in \mathcal{M}_j \cup \{\phi_j\}$, its signature can be computed as $\mathrm{Sig}_{sk_n}(tx_m^\mathrm{ID})\leftarrow \mathrm{H}_1(tx_m)^{\varsigma_{n} sk_n}$.
Within a pool $\Im_j$, the multi-signature can be computed by aggregating all members' signatures \cite{Boneh2001Short}, i.e.,
\begin{align}
\mathrm{MulSig}(tx_m^\mathrm{ID})  \leftarrow \prod\limits_{n \in \mathcal{M}_j \cup \{\phi_j\}} {\mathrm{Sig}_{sk_n}(tx_m^\mathrm{ID})}.
\end{align}
The correctness of multi-signature can be verified by checking whether $e(\mathrm{MulSig}(msg), g_2^{-1}) \cdot e(H_1(msg), apk_j) = 1_{\mathbb{G}_3}$.

\begin{figure}[!t]
\centering
  \includegraphics[width=7.cm]{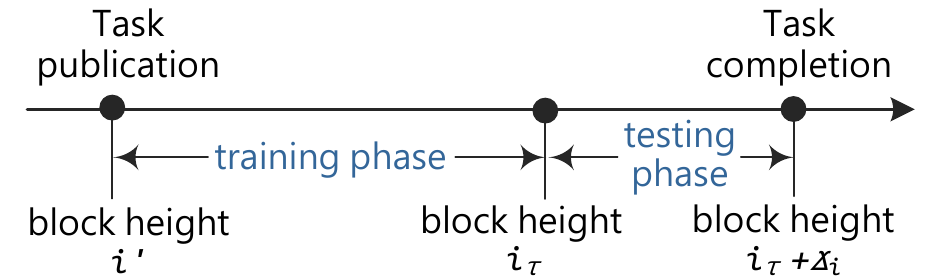}\\
  \caption{Illustration of training and testing phases of a FL task in PF-PoFL.}\label{fig:timephase}\vspace{-0.6mm}
\end{figure}

Once the training phase for a task $\tau$ has elapsed (i.e., current block height $i \ge i_\tau$), the task requester proceeds to submit the corresponding test dataset $\mathcal{D}_{\mathrm{test}}$ identified by the previously published hash value $\mathrm{dID}$ to the off-chain data storage.
The InterPlanetary File System (IPFS) \cite{benet2014ipfs} is adopted to serve as the distributed off-chain data store, where data is stored in the form of distributed files and is uniquely addressed through the hash pointer of the data.
After that, all involved pools proceed to upload their FL models to the IPFS identified by the hash value $\mathrm{mID}$. The training and testing phases for a FL task are depicted in Fig.~\ref{fig:timephase}.
To prevent malicious trainers from training a model using the released test data, the model transactions are regarded as invalid if they are generated during the task testing phase (i.e., the current block height is larger than the testing release block height).

\subsection{Model Ranking and Rewarding}\label{subsec:scheme4}
For each FL task to be completed, the more trained models associated with it, the higher chance for validators to earn more participation fees. Consequently, for all running tasks that are in the testing phase, it is reasonable to assume that validators are incentivized to reach consensus on the most urgent and profitable task (i.e., the task with the largest number of model transactions and shortest remaining test duration).
The validators in the set $\mathcal{V}$ independently execute the model ranking contract and block rewarding contract to compute the model ranking and complete financial settlements. The model ranking contract and block rewarding contract are shown in Algorithms~\ref{Algorithm0} and \ref{Algorithm1}, respectively.

\begin{algorithm}[t!]
   \caption{{Model Ranking Contract}}\label{Algorithm0}
    \begin{algorithmic}[1]
        \STATE \textbf{Input: }1) Task publication transaction ${tx}_\tau$; 2) Stable pooled structure $\Im^*$; 3) Model transactions ${tx}_m$ and payment transactions ${tx}_p$ related to the ${tx}_\tau$.
        \STATE \textbf{Output: }Model ranking ${mrList}_\tau$. 
        \STATE Extract $\{tx_\tau^\mathrm{ID},p_\tau,\xi_1, \psi,i_{\tau},\mathrm{dID}\}\leftarrow {tx}_\tau$.
        \STATE If ${verify}(tx_\tau)=true$, proceed to line 5, otherwise terminate.
        \STATE Add the model transactions ${tx}_m$ and payment transactions ${tx}_p$ related to the ${tx}_\tau$ into the local memory pool.
        \FOR{${tx}_m \in \mathcal{TX}_m^{\tau}$}
        \STATE Extract $\{\mathrm{mID},rs,\xi_2,{apk_j}\}\leftarrow {tx}_m$
        \STATE If $deposit_j\geq \xi_2$ \&\& $i_{m}\leq i_\tau$, proceed to line 9, otherwise terminate.
        \IF{the model $\Theta_{j,\tau}$ identified by $\mathrm{mID}$ is successfully downloaded from IPFS}
            \IF{the current height $i_\tau \le i \le i_\tau + \Delta_i$}
                \STATE Download task $\tau$'s test dataset $\mathcal{D}_{\mathrm{test}}$ identified by $\mathrm{dID}$.
                \STATE Evaluate performance score $ts_{j,\tau}$ on the test dataset $\mathcal{D}_{\mathrm{test}}$.
                \STATE Insert $\left<\mathrm{mID},ts_{j,\tau}\right>$ into the local model ranking ${mrList}_\tau$ depending on the performance metric.
            \ELSIF{the current height $i< i_\tau$}
                \STATE Wait until $i\geq i_\tau$, i.e., reaching the test phase.
            \ELSIF{the current height $i> i_\tau+ \Delta_i$}
                \STATE Terminate.
            \ENDIF
        \ELSE
            \STATE Set $ts_{j,\tau}=``unevaluated''$.
            \STATE Insert $\left<\mathrm{mID},``unevaluated''\right>$ into ${mrList}_\tau$.
        \ENDIF
        \ENDFOR
    \end{algorithmic}
\end{algorithm}

\emph{\textbf{Step 1:} Model ranking.}
As shown in Algorithm~\ref{Algorithm0}, each validator $v\in \mathcal{V}$ first validates the task publication transaction $tx_\tau$ by using $verify(tx_\tau)$ defined in Sect.~\ref{subsec:scheme2} (lines 3-4).
Then, validator $v$ adds all model transactions $tx_m$ and payment transactions $tx_p$ associated with the unfinished FL task $\tau\in\widetilde{\mathcal{T}_u}$ into its memory pool (line 5).
For each model transaction $tx_m$ issued by pool $\Im_j$, validator $v$ verifies (i) whether the deposit $deposit_j$ is no less than the participation fee $\xi_2$, and (ii) whether the issuing time (measured by block height) of transaction $tx_m$ (i.e., $i_m$) is no larger than the testing release block height $i_{\tau}$ (line 8).
For all valid model transactions, validator $v$ downloads the corresponding FL models identified by the hash pointer $\mathrm{mID}$ from the off-chain IPFS (line 9).

Once the task $\tau$ elapses its training phase and enters the testing phase (i.e., $i_\tau \le i \le i_\tau + \Delta_i$), validator $v$ downloads the test dataset $\mathcal{D}_{\mathrm{test}}$ identified by the previously published hash pointer $\mathrm{dID}$ from the IPFS and evaluates all the downloaded models (line 11). Here, $\Delta_i$ is the time duration (measured by block height) of the testing phase. For each model $\Theta_{j,\tau}$, it computes the performance score $ts_{j,\tau}$ on the test dataset $\mathcal{D}_{\mathrm{test}}$ (line 12). Based on the performance scores, each validator creates a candidate model ranking $mrList_\tau$, i.e., an ordered list of model-score pairs on the testing data, which is sorted either in descending or ascending order depending on the specific metrics $\psi$ (line 13).
Due to network latency, models that cannot be timely evaluated during the time limit $\Delta_i$ obtain a special ``unevaluated'' score in $mrList_\tau$ (lines 20--21).

\begin{algorithm}[t!]
   \caption{{Block Rewarding Contract}}\label{Algorithm1}
    \begin{algorithmic}[1]
        \STATE \textbf{Input: }1) Task publication transaction ${tx}_\tau$; 2) Stable pooled structure $\Im^*$ and curator $\phi_j,\forall \Im_j\in \Im^*$; 3) Consensus ranking ${mrList}_\tau$.
        \STATE \textbf{Output: }Financial rewarding transactions and node credits.
        \STATE Transfer $p_\tau\rightarrow wa_{j^*}$ from $esAdd$ to the pool $\Im_{j^*}$ that ranks best in ${mrList}_\tau$.
        \STATE Transfer $\xi_1 \!\cdot\! \frac{1}{\left|\mathcal{V}' \right|} + \xi_2 \cdot \frac{\left|\Im^*\backslash\mathcal{W}_\tau \right|}{\left|\mathcal{V}' \right|} \rightarrow wa_{v}$ from $esAdd$ to all non-faulty validators in the set $\mathcal{V}'$.
        \STATE Transfer $\xi_2 \rightarrow wa_{j}$ from $esAdd$ to all pools in $\mathcal{W}_\tau$ whose model is performed above the preset threshold.
        \STATE Update $cre_v \leftarrow cre_v + \chi_1, \forall v\in \mathcal{V}'$.
        \STATE Update $cre_m \leftarrow cre_m + \chi_2, \forall m\in \Im_{j^*}$.
        \STATE Update $cre_{m'} \leftarrow cre_{m'} + \chi_3, \forall m'\in \Im_{j}\subseteq \mathcal{W}_\tau,j\ne {j^*}$.
    \end{algorithmic}
\end{algorithm}

\emph{\textbf{Step 2:} Block rewarding.}
After reaching consensus on the model ranking for the selected task $\tau\in\widetilde{\mathcal{T}_u}$, validators will independently carry out the block rewarding procedure to enforce the reward and punishment for blockchain nodes, as shown in Algorithm~\ref{Algorithm1}. Specifically, the following financial transactions for rewarding and punishment will be added to the candidate block built by the validator.

1) A payment transaction transferring the task reward $p_{\tau}$ escrowed by the escrow address $esAdd$ to the wallet of the pool $\Im_{j^*}$ that owns the top-performing model in the consensus ranking $mrList_\tau$ (line 3). 

2) A payment transaction transferring the task hosting fee $\xi_1$ from $esAdd$ to all honest validators in the set $\mathcal{V}'$  (line 4). 
The fee $\xi_1$ is evenly distributed among all honest validators in $\mathcal{V}'$.

3) A payment transaction redeeming the participation fee $\xi_2$ to all pools in $\mathcal{W}_\tau$ whose models are performed above the preset threshold such as first quartile (line 5).

4) A payment transaction transferring the participation fee $\xi_2$ of all pools in $\Im^*\backslash\mathcal{W}_\tau$ whose models are either invalid or performed below the preset threshold to all honest validators (line 4). The total fees $\xi_2 \cdot {\left|\Im^*\backslash\mathcal{W}_\tau \right|}$ are evenly divided.

Apart from the economic rewards, the credits (as a virtual token) can offer incentives for blockchain nodes. The credit values of blockchain nodes can reflect their honest active participation statuses \cite{9343769}. After reaching consensus, the current credit value of the honest validators, members in the winning pool $\Im_{j^*}$, and trainers in $\mathcal{W}_\tau$ will be increased by $\chi_1$, $\chi_2$, and $\chi_3$, respectively (lines 6--8). Here, $\chi_k>0\,(k=1,2,3)$.

\subsection{Block Building and Commit}\label{subsec:scheme5}
A validator committee $\mathcal{V}$ is responsible for proposing and adding a new block to the blockchain by carrying out the proposed credit-based Algorand Byzantine agreement (BA) protocol with the following steps.

\emph{\textbf{Step 1:} Validator committee formation.} In conventional Algorand BA protocol \cite{Gilad2017Algorand}, nodes are weighted based on their account balance which can be only used for cryptocurrency applications. Instead, in our work, each node is weighted based on its owned credit value.
Thereby, our work is also applicable to non-cryptocurrency applications and can encourage nodes' participation willingness and honesty.
Specifically, members of the validator committee are randomly chosen based on their weights via a cryptographic sortition approach implemented by a verifiable random function (VRF).

A simple implementation of VRF using digital signature and hash function is adopted. At the beginning stage $s=1$, each node $n\in \mathrm{PK}(i-k)\subseteq\mathcal{N}$ independently validates whether it is a member of committee $\mathcal{V}_{i,s}$ by checking
\begin{align}\label{leaderselection}
.\mathrm{Hash}\left(\mathrm{Sig}_{sk_n}\left(i,s,seed_{i} \right) \right) \leq \frac{cre_{n}^i}{C} \!\cdot\! \frac{|\mathcal{V}_{i,s}|}{\#\mathrm{PK}(i\!-\!k)}.
\end{align}
In Eq. (\ref{leaderselection}), $seed_{i}$ is a public random seed for height $i$ which is included in the previous block with height $i-1$. ${cre_{n}^i}$ is the current credit value of node $n$. $C=\sum_{n\in \mathcal{N}}^{{cre_{n}^i}}$ is the total credit of blockchain nodes. $\#\mathrm{PK}(i-k)$ represents the number of public keys involved in consensus process between heights $i-k$ and $i$. The signature $\sigma_{n}^{i,s}=\mathrm{Sig}_{sk_n}\left(i,s,seed_{i} \right)$ means the credential of node $n$ to prove its role as a validator in $\mathcal{V}_{i,s}$ by revealing its public key $pk_n$. $\mathrm{Hash}\left(\sigma_{n}^{i,s}\right)$ is a random 256-bit long string uniquely determined by ${sk_n}$ and $i$, which is indistinguishable from random to any node that does not know the private key ${sk_n}$.
The symbol ``.'' indicates the transformation of hash value into a decimal in $(0,1]$.

\emph{\textbf{Step 2:} Candidate block building and propagation}.
All members in $\mathcal{V}_{i,1}$ are responsible for building candidate blocks and disseminating them to the whole network for verification.
The validator $v\in\mathcal{V}_{i,1}$ validates each transaction (i.e., $tx_\tau$, $tx_p$, and $tx_m$) in its memory pool that has not been included in the blockchain and is associated with an unfinished FL task $\tau\in\widetilde{\mathcal{T}_u}$. After that, it builds a candidate ranking $mrList_v$ according to the model ranking contract in Algorithm~\ref{Algorithm0}.
Then it orders all valid ones by timestamps, compresses them into a Merkle tree, and constructs a candidate block $\mathcal{B}_v^{ca}$.

\emph{\textbf{Step 3:} Mutual verification via multi-stage voting}.
The members in committee $\mathcal{V}_{i,s}$ are replaced in each consensus stage $s$ at height $i$ to prevent being targeted by adversaries. At stage $s>1$ of height $i$, $n\in \mathrm{PK}(i-k)$ is a validator in $\mathcal{V}_{i,s}$, if formula (\ref{leaderselection}) holds.
To mitigate block propagation delay, validators in $\mathcal{V}_{i,s}$ independently vote for hashes of candidate blocks, instead of the entire block proposal. The final agreement can be reached via the following two phases.

\underline{Phase 1}. All validators in $\mathcal{V}_{i,2}$ runs the $(N,f)$-graded consensus (GC) protocol \cite{chen2016algorand} to determine a value-grade pair $(\mu_v,\rho_v)$. 
Here, $f< N$ is the number of malicious or Byzantine nodes in PF-PoFL.
Specifically, each validator $v\in\mathcal{V}_{i,2}$ delivers its signed vote message $\mu_v=\mathrm{voteMsg}_v$ to other validators for mutual verification via the gossiping protocol, i.e.,
\begin{align}\label{votemsg}
\mathrm{voteMsg}_v = \left<\mathrm{Hash}(\mathcal{B}_{v'}^{ca}),i,s,seed_i,T_{\mathrm{stamp}},\right.\nonumber\\
\left.pk_v,\sigma_{n}^{i,s},\mathrm{Sig}_{sk_v}(\mathrm{voteMsg}_v) \right>,
\end{align}
where $\mathcal{B}_{v'}^{ca}$ is its voted candidate block.

For each validator $v\in\mathcal{V}_{i,2}$, if and only if $\#_v^2(\mathrm{voteMsg})\geq 2f+1$, it sends the vote message $\mathrm{voteMsg}$ to other validators. Besides, it sets the value-grade pair as $(\mu_v=\mathrm{voteMsg},\rho_v=2)$. If $\#_v^2(\mathrm{voteMsg})\geq f+1$, it sets $(\mu_v=\mathrm{voteMsg},\rho_v=1)$. Otherwise, it sets $(\mu_v=\bot,\rho_v=0)$.
Here, $\#_v^s(msg)$ means the number of validators from which $v$ has received the message $msg$ in stage $s$.

\underline{Phase 2}. Each validator in $\mathcal{V}_{i,s} (s>2)$ executes the improved binary BA (BBA*) protocol \cite{Gilad2017Algorand} (with initial input 0 if $\rho_v = 2$, and 1 otherwise) to determine the final consensus block and model ranking. If the result of BBA* is $out_v=0$, then the network reaches consensus on a candidate block voted by $\mu_v$. Otherwise, an empty block is recognized as the consensus block.

\begin{figure}[!t]
\centering\setlength{\abovecaptionskip}{-0.cm}
  \includegraphics[width=9.1cm]{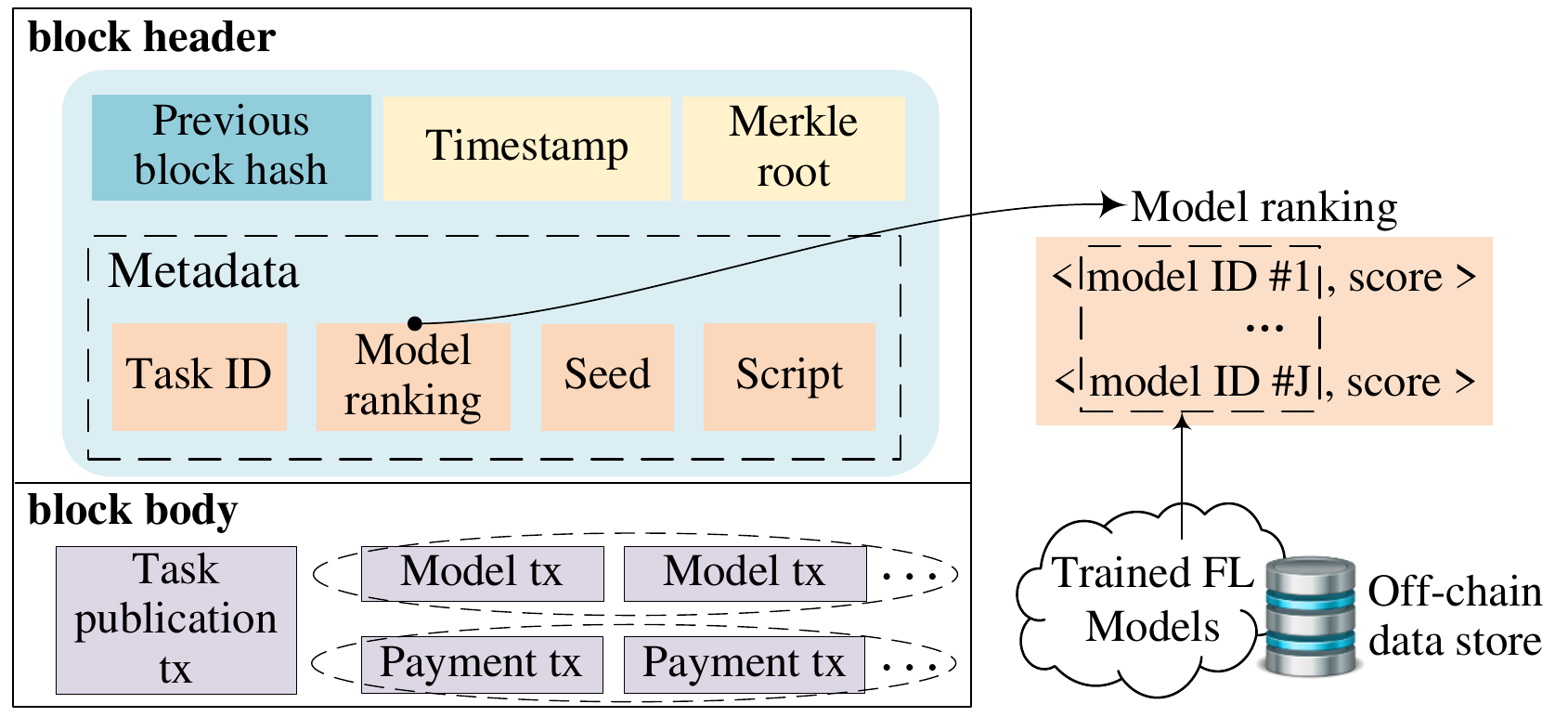}\\
  \caption{Structure of a block in PF-PoFL.}\label{fig:blockstructure}\vspace{-0.1cm}
\end{figure}

\emph{\textbf{Step 4:} Adding the consensus block}.
After reaching consensus, the newly built block $\mathcal{B}_{i}$ is successfully added into the blockchain which is linearly linked to the previous block with height $i-1$ via a hash pointer.
Fig.~\ref{fig:blockstructure} illustrates the structure of a block in our PF-PoFL.
The metadata $\mathrm{meta}_i$ in the newly built block $\mathcal{B}_{i}$ contains the identity of the completed FL task $\tau$, the consensus ranking $mrList_\tau$, the random seed $seed_i$ for VRF, and a script $Script$ that aggregates the signed votes of validators during the multi-stage voting for model ranking and consensus building. The script $Script$ for each block is agreed upon by the consensus process, which allows new users to catch up with the validation process of the block by processing these votes.
Besides, the $Script$ (as a certificate) allows any user to prove the safety of a block and efficiently verify the model ranking and credit updating process.

\subsection{Security and Privacy Analysis}\label{subsec:scheme6}
In this subsection, we give the privacy and security analysis of our PF-PoFL scheme.
To analyze the query sensitivity of the UDP mechanism in Eqs. (\ref{eq:scaledupdate})--(\ref{eq:globalagg}), an estimator $\hat{f}$ is defined as $\hat{f}(\Im_j)=\frac{1}{{|\Im_j|}} {\sum\nolimits_{m\in\Im_j}{{\Delta\Theta _{m}}}}$ in every communication round.
For privacy protection, the sensitivity of the query function $\hat{f}$, i.e., $\mathcal{S}(\hat{f})=\max_{\Im_j,m}||\hat{f}(\Im_j\cup\{m\})-\hat{f}(\Im_j) ||_2 $ should be controlled. Theorem~\ref{theorem0} analyzes the sensitivity bound of $\mathcal{S}(\hat{f})$.

\begin{theorem}\label{theorem0}
If ${||\Delta\tilde{\Theta} _{m}}||_2 \leq A$ holds $\forall m\in \Im_j$, the sensitivity of $\hat{f}$ is
bounded as $\mathcal{S}(\hat{f})\leq \frac{2A}{{|\Im_j|}}$.
\end{theorem}

\begin{IEEEproof}
For any $ m\in \Im_j$, as ${||\Delta\tilde{\Theta} _{m}}||_2 \leq A$, we have
\begin{align}
&||\hat{f}\left( \Im _j\cup \{m\} \right) -\hat{f}\left( \Im _j \right) ||\nonumber\\
&=\left|\left| \frac{\sum_{_{n\in \Im _j\cup \{m\}}}{\Delta \Theta _n}}{|\Im _j\cup \{m\}|}-\frac{\sum_{_{n\in \Im _j}}{\Delta \Theta _n}}{|\Im _j|} \right|\right| \nonumber\\
&=\left|\left| \frac{\Delta \Theta _m}{|\Im _j|+1}-\frac{\sum_{_{n\in \Im _j}}{\Delta \Theta _n}}{|\Im _j|\left( |\Im _j|+1 \right)} \right|\right| \\
&\le \left|\left| \frac{\Delta \Theta _m}{|\Im _j|+1} \right|\right| +\left|\left| \frac{\sum_{_{n\in \Im _j}}{\Delta \Theta _n}}{|\Im _j|} \right|\right| \frac{1}{|\Im _j|+1} \le \frac{2A}{|\Im _j|}.\nonumber
\end{align}
Theorem~\ref{theorem0} is proved.
\end{IEEEproof}

\begin{theorem}\label{theorem1}
Given the number of communication rounds $\Psi_{j,\tau}^{\mathrm{global}}$ and the sample ratio $\lambda_s$, $\forall \epsilon < 2(\lambda_s)^2\log(1/\lambda_s)\Psi_{j,\tau}^{\mathrm{global}}$ and $\forall \delta>0$, the proposed mechanism satisfies $(\epsilon,\delta$)-UDP, if
\begin{align}
\sigma \geq \frac{2\lambda_s\sqrt{\Psi_{j,\tau}^{\mathrm{global}}\log(1/\delta)}}{\epsilon}.
\end{align}
\end{theorem}

\begin{IEEEproof}
According to the Lemma 3 of \cite{Abadi2016Deep}, given $\sigma>1$ and $\lambda_s< \frac{1}{16\sigma}$, for any positive integer $\gamma\leq \sigma^2 \ln(1/{\lambda_s\sigma})$, the moment of UDP mechanism $\mathcal{F}$ satisfies
\begin{align}
\alpha_{\mathcal{F}}(\gamma)\leq \frac{(\lambda_s)^2\gamma(\gamma+1)} {(1-\lambda_s)\sigma^2}+\mathcal{O}\Big(\Big(\frac{\lambda_s \gamma}{\sigma}\Big)^3\Big),
\end{align}
which is bounded by $\alpha_{\mathcal{F}}(\gamma)\leq \frac{(\lambda_s)^2\gamma(\gamma+1)} {(1-\lambda_s)\sigma^2}$. Besides, based on the Theorem 2 of \cite{Abadi2016Deep}, to be $(\epsilon,\delta$)-UDP, it suffices that
\begin{align}\label{twocon}
\frac{\Psi_{j,\tau}^{\mathrm{global}}(\lambda_s)^2\gamma^2}{\sigma^2}\leq \frac{\gamma\epsilon}{2}, \ \mathrm{and}\ \delta \geq \exp\left(-\frac{\gamma\epsilon}{2}\right).
\end{align}
According to formula (\ref{twocon}), we can derive that
\begin{align}\label{sigma}
\sigma \geq \frac{\lambda_s\sqrt{2\Psi_{j,\tau}^{\mathrm{global}}\gamma\epsilon}}{\epsilon}\geq \frac{2\lambda_s\sqrt{\Psi_{j,\tau}^{\mathrm{global}}\log(1/\delta)}}{\epsilon}.
\end{align}
As $\sigma>1$ and $\lambda_s\in(0,1]$, we have $\ln(1/{\lambda_s\sigma})< \ln(1/{\lambda_s})$. According to $\delta \geq \exp\left(-\frac{\gamma\epsilon}{2}\right)$, we have $\gamma \epsilon \geq 2\log \left( 1/\delta \right)$.
By transforming $\gamma\leq \sigma^2 \ln(1/{\lambda_s\sigma})$, we have
\begin{align}
\epsilon&\leq \frac{4\left( \lambda _s \right) ^2\Psi _{j,\tau}^{\text{global}}\log \left( 1/\delta \right) \log \left( 1/\lambda _s\sigma \right)}{\gamma \epsilon} \\
&< 2(\lambda_s)^2 \Psi_{j,\tau}^{\mathrm{global}}\log(1/\lambda_s\sigma)< 2(\lambda_s)^2 \Psi_{j,\tau}^{\mathrm{global}}\log(1/\lambda_s).\nonumber
\end{align}
Theorem~\ref{theorem1} is proved.
\end{IEEEproof}

In the blockchain, the reasonable security assumption is adopted, where the fraction of the credit value held by malicious blockchain nodes is assumed to be less than $1/3$.
In the following, we analyze the finality, consistency, and efficiency of our PF-PoFL system, as well as security analysis in defending against adversaries defined in Sect. \ref{subsec:threatmodel}.
\begin{itemize}
  \item \emph{Finality and consistency}. According to \cite{Gilad2017Algorand,chen2016algorand}, under the above majority honest assumption, our PF-PoFL system is resilient up to $1/3$ credit values held by malicious or Byzantine nodes, and all honest nodes can reach an agreement on the same block with overwhelming probability. It ensures the finality and consistency of the blockchain, as well as the prevention of blockchain forks and double-spending threats.
  \item \emph{Consensus efficiency}. Under the \emph{strong synchrony} assumption (i.e., messages sent by most honest nodes can be received by most other honest nodes within a known time bound), the BBA* protocol is proved to reach agreement within one stage, thereby ensuring the high efficiency of consensus protocol in common situations.
  \item \emph{Defense of spoofing attack}. In the model ranking smart contract, a model transaction will not be included in the candidate block by honest validators if it is published within the testing phase of the task. Hence, trainers do not have incentives to cheat by training a perfect model on the test data. Besides, as only the hashes of trained FL models are released in federated mining process, adversaries cannot plagiarize the FL models trained by other pools. Besides, the payment delivery to members of the winning pool is executed automatically via the block rewarding smart contract, the refusal-to-pay threat can be avoided. Thereby, PF-PoFL can resist spoofing attacks.
  \item \emph{Defense of Sybil attack}. By introducing the participation fee, trainers/pools have to pay a participation fee before their trained models are ranked, which increases the cost for nodes to carry out Sybil attacks. Besides, as the committee members are randomly formed via VRF based on their credit values, Sybil attacks to the committee selection process can be prevented, as well as ensuring fairness for blockchain nodes. Thereby, PF-PoFL can mitigate Sybil attacks.
  \item \emph{Defense of the central platform}. PF-PoFL offers a fully decentralized solution for the life-cycle process including task outsourcing, model training, model ranking, and reward sharing. In PF-PoFL, as the third-party platform is removed, potential risks it brings including SPoF, collusion, and sensitive information leakage can be avoided.
  \item \emph{Privacy leakage prevention}. According to Theorem~\ref{theorem1}, PF-PoFL ensures UDP for miners to prevent user-level privacy leakage during federated mining. Besides, as the committee selection is pseudo-random and non-interactive and the committee members are replaced in every stage, adversaries cannot target a committee member until that validator starts participation. Thereby, PF-PoFL also ensures membership unpredictability for validators to prevent from being targeted.
\end{itemize}

\section{Dynamic Optimized Pool Structure Formation}\label{sec:COALITION}
In this section, we first analyze the federation utility of each pool, then we present a stable pool formation algorithm based on the federation formation game, followed by the property analysis of the proposed algorithm.

\subsection{Federation Utility Function}\label{subsec:coalition1}
The federation utility of pool $\Im_j$ is determined by the accuracy loss of the trained global model and the federation cost.
First, we characterize the expected accuracy loss of the global model under federated mining in Sect.~\ref{subsec:scheme3}.

\emph{1) Non-IID effect}. The non-IID data can affect the accuracy of the global model under FL. We adopt the widely used average earth mover's distance (EMD) to measure the heterogeneity of data distribution among diverse miners \cite{zhao2018federated}. In a typical classification problem with $Y$ classes defined over a label space $\mathcal{Y} = \{1, \cdots, Y\}$ and a compact space $\mathcal{X}$. Let $\mathrm{Pr}_m(y=k)$ denote the proportion of miner $m$'s local training samples with label $k\in\mathcal{Y}$, and $\mathrm{Pr}(y=k)$ be the proportion of training samples with label $k\in\mathcal{Y}$ in all miners' training data. Thereby, $\mathrm{Pr}(y),\forall y \in \mathcal{Y}$ denotes the data distribution for each data sample $\{\mathbf{x}, y\}$ distributed over $\mathcal{X} \times \mathcal{Y}$. The EMD of miner $m\in \mathcal{M}_{\tau}$ is computed as
\begin{align}
\Xi_m = \sum_{k=1}^{Y}{||\mathrm{Pr}_m(y=k) - \mathrm{Pr}(y=k)||}.
\end{align}
The average EMD in pool $\Im_j$ is computed as the weighted sum of all miners' EMDs, which is expressed as
\begin{align}\label{nonIID}
\overline{\Xi}_j = \sum\nolimits_{m=1}^{|\Im_j|}{\frac{s_m}{\sum_{m=1}^{|\Im_j|}{s_m}}}\Xi_m,
\end{align}
where $s_m$ is miner $m$'s local data size, and ${|\Im_j|}$ is the number of members in pool $\Im_j$.

\emph{2) Latency analysis}. In a global communication round $k$, the total latency consists of (i) local training latency $D_{\mathrm{tl}}^k$ in local model training and (ii) network latency $D_{\mathrm{nl}}^k$ for uploading the trained local model and receiving the latest global model using wired/wireless networks.
According to \cite{Konecn2016FederatedOD}, we have $D_{\mathrm{nl}}^k\gg D_{\mathrm{tl}}^k$, thereby $D_{\mathrm{tl}}^k$ can be negligible. Here, the maximum waiting time (i.e., $D_j^{\max}$) of pool $\Im_j$ in a global communication round is defined as
\begin{align}
D_j^{\max}=\beta \overline{D}_{\mathrm{nl}}=\beta \sum_{m=1}^{|\Im_j|}{\overline{D}_{\mathrm{nl},m}},
\end{align}
where $\beta>0$ is an adjustment factor to control the ratio of participants in a global communication round. $\overline{D}_{\mathrm{nl}}$ is the average network latency of all miners in the pool $\Im_j$. ${\overline{D}_{\mathrm{nl},m}}$ is the expected network latency of miner $m\in\Im_j$.

\emph{3) Number of training samples}. In a global communication round, the total number of data samples used for model training in pool $\Im_j$ is denoted as $S_j = \sum_{m=1}^{|\Im_j|}{\alpha_m s_m}$. Here, $\alpha_m=\{0,1\}$ is a binary variable, i.e.,
\begin{align}
\alpha_m =
\begin{cases}
1, & \mathrm{if}\ {\overline{D}_{\mathrm{nl},m}}\le D_j^{\max},\\
0, & \mathrm{otherwise}.
\end{cases}
\end{align}
It means that only miners with ${\overline{D}_{\mathrm{nl},m}}\le D_j^{\max}$ are admitted to join the model training.

\emph{4) Total global communication rounds}. The number of global communication rounds of pool $\Im_j$ can be computed as $\Psi_{j,\tau}^{\mathrm{global}}=\lfloor T_\tau^{\mathrm{train}}/D_j^{\max}\rfloor$. Here, $T_\tau^{\mathrm{train}}$ is the total training time duration (measured by block height) of a FL task, as shown in Fig.~\ref{fig:timephase}.

\emph{5) Expected accuracy loss}. The accuracy loss of the global model after $\Psi_{j,\tau}^{\mathrm{global}}$ rounds can be measured by the prediction loss between the optimal parameter $\Theta^*$ and the parameter $\Theta(\Psi_{j,\tau}^{\mathrm{global}})$, i.e., $\mathcal{L}(\Theta(\Psi_{j,\tau}^{\mathrm{global}})) - \mathcal{L}(\Theta^*)$. According to \cite{9252911,li2014efficient}, when adopting mini-batch SGD for local training, the expected accuracy loss $\Pi_j$ under non-IID case is bounded by $\mathcal{O}\Big(\mathcal{E}(\overline{\Xi}_j)\Big(\frac{1}{\sqrt{S_j \Psi_{j,\tau}^{\mathrm{global}}}} + \frac{1}{\Psi_{j,\tau}^{\mathrm{global}}}\Big)\Big)$. Here, $\mathcal{E}(\overline{\Xi}_j)$ is the relative accuracy loss function. Obviously, the expected accuracy loss decreases with the increase of the number of global iterations $\Psi_{j,\tau}^{\mathrm{global}}$ and the number of training samples $S_j$.

Apart from the federated mining strategy, miners can also employ the solo mining strategy, in which the miner $m$ chooses to work alone and forms a singleton, i.e., $\Im_j=\{m\}$. Under this case, the expected accuracy loss bound can be approximated as $\mathcal{O}\Big(\mathcal{E}(\Xi_m)\Big(\frac{1}{\sqrt{s_m \Psi_{\max,\tau}^{\mathrm{local}} }} + \frac{1}{\Psi_{\max,\tau}^{\mathrm{local}}}\Big)\Big)$, where $\Psi_{\max,\tau}^{\mathrm{local}}$ is the maximum number of local epoches in training task $\tau$ to avoid overfitting. To summarize, the explicit form of the expected accuracy loss is expressed as
\begin{align}\label{accuracyloss}
\Pi_j \!=\!
\begin{cases}
\mathcal{E}(\overline{\Xi}_j)\Big(\frac{1}{\sqrt{S_j \Psi_{j,\tau}^{\mathrm{global}}}} + \frac{1}{\Psi_{j,\tau}^{\mathrm{global}}}\Big), & \mathrm{if}\ {|\Im_j|}> 1,\\
\mathcal{E}(\Xi_m)\Big(\frac{1}{\sqrt{s_m \Psi_{\max,\tau}^{\mathrm{local}}}} + \frac{1}{\Psi_{\max,\tau}^{\mathrm{local}}}\Big), & \mathrm{if}\ {|\Im_j|}=1.
\end{cases}
\end{align}

\emph{6) Federation satisfaction function}. The federation satisfaction function $\mathcal{S}(\Pi_j)$ measures the satisfaction that the pool $\Im_j$ receives in the expected accuracy loss $\Pi_j$. Generally, the lower the accuracy loss, the higher the federation satisfaction. Besides, it becomes harder to further reduce the accuracy loss when the loss is smaller, indicating that the faster the drop rate of accuracy loss, the higher the obtained federation satisfaction.
Thereby, $\mathcal{S}(\Pi_j)$ should satisfy $\mathcal{S}(\Pi_j)\geq 0$, $\frac{\mathrm{d} \mathcal{S}(\Pi_j)}{\mathrm{d}\Pi_j}<0$, and $\frac{\mathrm{d}^2 \mathcal{S}(\Pi_j)}{\mathrm{d}\Pi_j^2}\ge 0$. A well-suited satisfaction function satisfying these requirements can be the exponential decay function \cite{6081879} given by 
\begin{align}
\mathcal{S}(\Pi_j) = \gamma_s\exp(-\gamma_d \cdot \Pi_j),
\end{align}
where $\gamma_s>0$ is the satisfaction parameter, and $\gamma_d>0$ is the decay parameter which controls the decay speed. A larger $\gamma_d$ indicates a faster decay of satisfaction.

Next, we analyze the federation cost in pool formulation.

\emph{7) Federation cost function}. The miners in pool $\Im_j$ need to frequently synchronize the latest global model from the curator. According to \cite{6550882,5678781}, the federation cost $\mathcal{C}(\Im_j)$ can be measured by the communication cost which varies linearly with the federation size ${|\Im_j|}$. We have
\begin{align}
\mathcal{C}(\Im_j) =
\begin{cases}
\lambda_c{|\Im_j|}, & \mathrm{if}\ {|\Im_j|}> 1,\\
0, & \mathrm{if}\ {|\Im_j|}=1,
\end{cases}
\end{align}
where $\lambda_c>0$ is a scaling factor.

Finally, the federation utility of pool $\Im_j$ can be obtained as the difference between the federation satisfaction and the federation cost, i.e.,
\begin{align}
&~~\mathcal{U}(\Im_j) = \mathcal{S}(\Pi_j) - \mathcal{C}(\Im_j) \nonumber \\
&\!=\!
\begin{cases}
\gamma_s\exp\left(-\gamma_d  \mathcal{E}(\overline{\Xi}_j)\Big(\frac{1}{\sqrt{\sum_{m=1}^{|\Im_j|}{\alpha_m s_m} \big\lfloor \frac{T_\tau^{\mathrm{train}}}{D_j^{\max}}\big\rfloor }} \!+\! {\Big\lfloor \frac{T_\tau^{\mathrm{train}}}{D_j^{\max}}\Big\rfloor}^{\!-\!1}\Big)\right) \\ ~~~~~~~~~~~~~~~~~~~~~~~~~~~~~~~~~~~~~~~~ -\lambda_c{|\Im_j|},~ \mathrm{if}\ {|\Im_j|}> 1,\\
\gamma_s\exp\left(-\gamma_d \mathcal{E}(\Xi_j)\Big(\frac{1}{\sqrt{s_j \Psi_{\max,\tau}^{\mathrm{local}}}} + \frac{1}{\Psi_{\max,\tau}^{\mathrm{local}}}\Big) \right), \mathrm{if}\ {|\Im_j|}=1.
\end{cases}
\end{align}

Let $W(\Im)$ denote the social welfare of a partition $\Im$, i.e., the sum of federation utilities of disjoint pools in a partition $\Im$.
The target of PF-PoFL is to maximize the social welfare by forming the optimal partition structure $\Im^*$, where the optimization problem for task $\tau$ is formulated as:
\begin{align}\label{socialwelfare}
\mathbf{Problem:}\, \max W(\Im):=\sum\nolimits_{j=1}^{|\Im|}{\mathcal{U}(\Im_j)}.\nonumber \\
\mathrm{s.t.}\ {\Im_j}\cap{\Im_{j'}}=\emptyset, \forall j \ne j',  \cup_{j=1}^{{|\Im|}}{\Im_j}=\mathcal{M}_{\tau}.
\end{align}

\subsection{Stable Pool Formation Algorithm}\label{subsec:coalition2}
To solve the problem in (\ref{socialwelfare}), the pooled-mining formation process among miners is modeled as a federation formation game with transferable utility (FFG-TU), where distributed miners (as game players) tend to form various disjoint federations to maximize their individual profits \cite{9665214}.

\emph{\textbf{Definition 2} (FFG-TU game):}
For each uncompleted FL task $\tau$, a FFG-TU game for optimized pool structure is defined by a triple $(\mathcal{M}_{\tau}, \mathcal{U},\Im)$, where the specific formulation is shown as below.
\begin{itemize}
  \item \emph{Players:} The players of the game is the set of miners (i.e., $\mathcal{M}_{\tau}$) that participate FL task $\tau$.
  \item \emph{Transferable federation utility:} $\mathcal{U}(\Im_j)$ is the federation utility of each federation (or pool) $\Im_j\subseteq \mathcal{M}_{\tau}$, which can be apportioned arbitrarily among the members of $\Im_j$.
  \item \emph{Pooled structure:} A pooled structure (or a federation partition) is defined as the set $\Im = \{\Im_1,\cdots,\Im_J\}$ that partitions the miner set $\mathcal{M}_{\tau}$. Here, $\Im_j,j=1,\cdots,J$ are disjoint federations such that ${\Im_j}\cap{\Im_{j'}}=\emptyset$, $\forall j \ne j'$, and $\cup_{j=1}^{J}{\Im_j}=\mathcal{M}_{\tau}$.
  \item \emph{Strategy:} Each miner decides whether to work alone by adopting the \emph{solo mining} strategy or choose a federation to join to cooperatively train a FL model using the \emph{federated mining} strategy.
\end{itemize}

\emph{\textbf{Definition 3} (Switch operation):}
A switch operation $\Phi_{l,k}(m)$ is defined as the transfer of miner $m$ from the current pool $\Im_l\in \Im$ to another pool $\Im_{k}\in \Im\cup\{\emptyset\}$, mathematically,
\begin{align}
\Phi_{l,k}(m): \Im_l\triangleright \{\Im_l^-,\{m\}\} \ \mathrm{and}\ \{\Im_k,\{m\}\}\triangleright \Im_{k}^+.
\end{align}

\begin{remark}
The switch operation consists of two successive parts: the split operation (i.e., $\Im_l\triangleright \{\Im_l^-,\{m\}\}$) and the merge operation (i.e., $\{\Im_k,\{m\}\}\triangleright \Im_{k}^+$), where $\Im_l^-=\Im_l\backslash\{m\}$ and $\Im_l^+=\Im_k\cup\{m\}$.
In the case that $\Im_{l}=\{m\}$, $\Phi_{l,k}(m)$ means the merge of $\Im_l$ with $\Im_k$, causing the number of federations decrease by one.
In the case that $\Im_{k}=\{\emptyset\}$, $\Phi_{l,k}(m)$ means the formation of a new singleton $\Im_{k}=\{m\}$, causing the number of federations increase by one.
In the case that $k=l$, it means miner $n$ stays in the current federation and the number of federations remains unchanged.
\end{remark}

\emph{\textbf{Definition 4} (Switch gain):} The switch gain $\Omega_m(\Phi_{l,k})$ related to the switch operation $\Phi_{l,k}(m)$ is defined as:
\begin{align}
&\Omega_m(\Phi_{l,k}) := v_m(\Im_k\cup\{m\}) - v_m(\Im_l)\nonumber \\
&= [\mathcal{U}(\Im_k\cup\{m\}) - \mathcal{U}(\Im_k)] - [\mathcal{U}(\Im_l) - \mathcal{U}(\Im_l \backslash \{m\})].
\end{align}
Here, $v_m(\Im_k\cup\{m\})$ and $v_m(\Im_l)$ mean the extra utility value that miner $m$ brings to the pools $\Im_k\cup\{m\}$ and $\Im_l$, respectively.

\begin{remark}
Specially, we define $\Omega_m(\Phi_{l,k})=0$ when $k=l$. Besides, when $\Im_k=\{\emptyset\}$, we have $\Omega_m(\Phi_{l,\emptyset}) = \mathcal{U}(\{m\}) - v_m(\Im_l)$.
The transferable switch gain in the pool switch operation accords with the transferable utility of our proposed federation formation game. Besides, the federation gain of any switch operation only relies on the utilities of involved pools, rather than the manner that federation utilities are shared among their members.
\end{remark}

\emph{\textbf{Definition 5} (Preference order):} For any miner $m\in \mathcal{M}_\tau$, the preference order $\succeq$ is defined as a complete and transitive binary relation between two switch
operations $\Phi_{l,k}(m)$ and $\Phi_{l',k'}(m)$ such that:
\begin{align}
\Phi_{l,k}(m) \succeq \Phi_{l',k'}(m) \Leftrightarrow \Omega_m(\Phi_{l,k}) \geq \Omega_m(\Phi_{l',k'}).
\end{align}
Similarly, for the strict preference order $\succ$, we also have $\Phi_{l,k}(m) \succ \Phi_{l',k'}(m) \Leftrightarrow \Omega_m(\Phi_{l,k}) > \Omega_m(\Phi_{l',k'})$.

As miners can transfer from one federation to another, the federation partition (i.e., ${\Im}$) varies with time. The solution of the proposed FFG-TU game is a stable pooled structure $\Im^*$. In general, the stable partition outcome can be attained using exhaustive search methods, where the possible partition iterations increase exponentially with the number of involved miners \cite{5678781}.
In the following, we design a distributed stable pool formation algorithm with low computational complexity in Algorithm~\ref{Algorithm2} to obtain the optimal federation partition strategy of each player (i.e., miner $m$) in the game. The proposed algorithm consists of the following three phases. 

\begin{algorithm}[t!]
   \caption{\textbf{Distributed Pool Formation Algorithm}}\label{Algorithm2}
    \begin{algorithmic}[1]
        \STATE \textbf{Input: }$\mathcal{T}_u$, $\mathcal{M}_\tau$, $\Xi_m,{s_m}$, $\beta$, ${\overline{D}_{\mathrm{nl},m}}$, $T_\tau^{\mathrm{train}}$, ${\Psi_{m,\tau}^{\mathrm{local}}}$, $\gamma_s$, $\gamma_d$, $\lambda_c$.
        \STATE \textbf{Output: }Stable federation structure $\Im^*$.
        \STATE \textbf{Initialization: }Set $h=0$ and the initial pool structure $\Im^{(0)}$.
        \STATE \textbf{Repeat }
            \FOR{ $m \in \mathcal{M}_\tau$ }
                \STATE Calculate all candidate pools (including the empty set $\{\emptyset\}$) with positive switch gain, and record them into the set $\mathcal{C}_m^{\mathrm{pool}}$.
                \STATE Send transfer request to most preferred pool ${\Im_{j^*}^{(h)}}\in \mathcal{C}_m^{\mathrm{pool}}$ such that 
                \begin{align}\label{eq:}
                \Omega_m(\Phi_{l,j^*})\!\geq\! \Omega_m(\Phi_{l,j}), m\!\in\!{\Im_{l}^{(h)}},\forall {\Im_j^{(h)}} \in \mathcal{C}_m^{\mathrm{pool}},
                \end{align}
            where ${\Im_{l}^{(h)}}$ is the current federation of miner $m$ at $h$-th iteration.
            \ENDFOR
            \FOR{ ${\Im_j^{(h)}} \in  \Im^{(h)}$ }
            \STATE Record the candidate miners that send a transfer request into the set $\mathcal{C}_j^{\mathrm{miner}}$.
            \STATE Accept the most preferred miner $m^*\in{\Im_{l'}^{(h)}}$ through the admission rule:
                \begin{align}\label{eq:}
                \Omega_{m^*}(\Phi_{l',j})\!\geq\! \Omega_m(\Phi_{l,j}), m\!\in\!{\Im_{l}^{(h)}},\forall m \!\in\! \mathcal{C}_{j}^{\mathrm{miner}}.
                \end{align}
            \STATE Reject other candidate miners in the set $\mathcal{C}_{j}^{\mathrm{miner}}\backslash \left\{ {{m^*}} \right\}$.
            \STATE Do split operation, i.e., ${\Im _{l'}^{(h)}}  \triangleright\left\{ {{\Im _{l'}^{(h+1)}},\left\{ {{m^*}} \right\}} \right\}$, where ${\Im _{l'}^{(h+1)}} = {\Im _{l'}^{(h)}}\backslash \left\{ {{m^*}} \right\}$.
            \STATE Do merge operation, i.e., $\left\{ {{\Im _{j}^{(h)}},\left\{ {{m^*}} \right\}} \right\}\triangleright {\Im _{j}^{(h+1)}}$, where ${\Im _{j}^{(h+1)}} = {\Im _{j}^{(h)}} \cup \left\{ {{m^*}} \right\}$.
            \ENDFOR
        \STATE $h = h + 1$.
        \STATE Update the federation structure as ${\Im}^{(h)} = \{\Im_1^{(h)},\cdots,\Im_{J^{(h)}}^{(h)}\}$.
        \STATE \textbf{Until }no miner do switch operations, i.e., $\Phi_{l,l}(m)\succeq\Phi_{l,j}(m)$, $m\in \Im_l^{(h)}$, $j\ne l$, $\forall m \in \mathcal{M}_\tau$, $\forall \Im_j^{(h)} \in {\Im}^{(h)}\cup \{\emptyset\}$.
    \end{algorithmic}
\end{algorithm}

\underline{Phase 1: Pool initialization} (line 3). At the beginning ($h=0$), an initial partition ${\Im ^{(0)}}$ is generated according to the specific application. For example, the initial partition can be the result of the stable partition outcome for the previously finished FL task.

\underline{Phase 2: Switch strategy-making on miner's side}~(lines 5--8). Given the current federation partition ${\Im}^{(h)} = \{\Im_1^{(h)},\cdots,\Im_{J^{(h)}}^{(h)}\}$, each miner can opt three strategies: (i) stay in the current federation; (ii) split from the current federation and merge with any other non-empty federation; and (iii) split from the current federation and act alone. The first two strategies are federated mining strategies and the last one is the solo mining strategy. To formulate each miner's switch strategy, the switchable candidate pool is first defined as below.

\emph{\textbf{Definition 6} (Switchable candidate pool):} For each miner $m\in {\Im_l}$, its switchable candidate pool set is defined as $\mathcal{C}_{m}^{\mathrm{pool}}$, where $\Omega_m(\Phi_{l,k})>0, \forall \Im_k \in \mathcal{C}_{m}^{\mathrm{pool}}\subseteq \Im\cup\{\emptyset\}$. In other words, only the pool with positive switch gain can be added into the switchable candidate pool set.

\begin{remark}
If $\Omega_m(\Phi_{l,k})\leq 0, \forall \Im_k \in \Im\cup\{\emptyset\}$, there exists no available federation in $\Im$ nor the empty set to transfer for the miner $m$ (i.e., $\mathcal{C}_{m}^{\mathrm{pool}}=\emptyset$), indicating that the miner intends to stay in the current federation $\Im_{l}$ (i.e., $\Phi_{l,l}(m)$). Otherwise, the miner makes its strategy based on the following switch rule.
\end{remark}

\emph{\textbf{Definition 7} (Switch rule):} Given the switchable candidate pool set $\mathcal{C}_{m}^{\mathrm{pool}}\cup\{\emptyset\}$, each miner $m\in {\Im_l}$ sends transfer request to the candidate pool $\Im_{j^*}$ with the largest switch gain, i.e.,
\begin{align}
\Im_{j^*} = \operatorname{arg\,max} \Omega_m(\Phi_{l,j}), m \in {\Im_l}, \forall \Im_{j} \in \mathcal{C}_{m}^{\mathrm{pool}}\cup\{\emptyset\}, 
\end{align}
where ${\Im_l}$ is the current federation of miner $m$.

\begin{remark}
In the case $\operatorname{arg\,max} \big\{\Omega_m(\Phi_{l,j})\!:\!\forall \Im_{j} \in \mathcal{C}_{m}^{\mathrm{pool}}\cup\{\emptyset\} \big\}\!=\!\{\emptyset\}$, it means that the miner $m$ prefers the solo mining strategy and will split from the current federation (i.e., $\Phi_{l,\emptyset}(m)$). In the other case $\operatorname{arg\,max} \left\{\Omega_m(\Phi_{l,j})\!:\!\forall \Im_{j} \in \mathcal{C}_{m}^{\mathrm{pool}}\cup\{\emptyset\} \right\}=\Im_{j^*}$, it means that the miner $m$ prefers splitting from the current federation $\Im_{l}$ and merging with another federation $\Im_{j^*}$ (i.e., $\Phi_{l,j^*}(m)$).
\end{remark}

To summarize, $\forall \Im_{j} \in \mathcal{C}_{m}^{\mathrm{pool}}\cup\{\emptyset\}$, the optimal switch strategy of each miner $m\in\Im_{l}$ can be expressed as
\begin{align}\label{minerstrategy}
\Phi_{l,k}^*(m) =
\begin{cases}
\Phi_{l,l}(m), & \mathrm{if}\ \mathcal{C}_{m}^{\mathrm{pool}}=\emptyset,\\
\Phi_{l,\emptyset}(m), & \mathrm{if}\ \operatorname{arg\,max} \Omega_m(\Phi_{l,j})\!=\!\{\emptyset\},\\
\Phi_{l,j^*}(m), & \mathrm{if}\ \operatorname{arg\,max} \Omega_m(\Phi_{l,j})\!=\!\Im_{j^*}.
\end{cases}
\end{align}

\underline{Phase 3: Admission strategy-making on pool's side} (lines 9--17).
When multiple miners request to join the same pool $\Im_j$, the switch order can affect the federation utilities and the federation formation outcome. To capture the pool's preference for the transfer order, the admission rule is defined as below.

\emph{\textbf{Definition 8} (Admission rule):} For a set of candidate miners (i.e., $\mathcal{C}_{j}^{\mathrm{miner}}$) that send transfer request to the same pool $\Im_j$, only the candidate $m^*$ with the largest switch gain is permitted by the pool $\Im_j$, i.e.,
\begin{align}
{m^*} = \operatorname{arg\,max} \Omega_m(\Phi_{l,j}), m \in {\Im_l}, \forall m \in \mathcal{C}_{j}^{\mathrm{miner}},
\end{align}
and other candidate miners in the set $\mathcal{C}_{j}^{\mathrm{miner}}\backslash \left\{ {{m^*}} \right\}$ are rejected.

According to the admission rule, each pool $\Im_j\in\Im$ accepts the most preferred miner $m^*\in \Im_{l'}$ (in line 11) and rejects other candidates (in line 12). Correspondingly, the switch operation is performed between two pools $\Im_j,\Im_{l'}\in\Im$ consisting of the split operation (i.e., ${\Im _{l'}^{(h)}}  \triangleright\big\{ {{\Im _{l'}^{(h+1)}},\left\{ {{m^*}} \right\}} \big\}$) and the merge operation (i.e., $\big\{ {{\Im _{j}^{(h)}},\left\{ {{m^*}} \right\}} \big\}\triangleright {\Im _{j}^{(h+1)}}$), where ${\Im _{l'}^{(h+1)}} = {\Im _{l'}^{(h)}}\backslash \left\{ {{m^*}} \right\}$ and ${\Im _{j}^{(h+1)}} = {\Im _{j}^{(h)}} \cup \left\{ {{m^*}} \right\}$. Then, the federation structure is updated as ${\Im}^{(h)}\rightarrow{\Im}^{(h+1)}$ (in line 17).

The above phases 2-3 are repeated until no miner $m \in \Im_l^*\subseteq \mathcal{M}_\tau$ in the final pooled structure $\Im^*$ intends to execute switch operations to increase the overall gain by transferring to another pool $\Im_j^* \in {\Im}^{*}\cup \{\emptyset\}$ with $j\ne l$, i.e.,
$\Phi_{l,l}(m)\succeq\Phi_{l,j}(m)$.

\subsection{Property Analysis}\label{subsec:coalition3}
Given any initial partition $\Im^{(0)}$, the federation formation procedure can be formulated as a sequence of switch operations, i.e.,
\begin{align}\label{switchsequence}
\{{\Im ^{(0)}} \to  \cdots  \to {\Im ^{(h)}} \to  \cdots  \to {\Im ^{(H)}}={\Im}^{*}\},
\end{align}
where $H$ is total number of transformations.

\begin{lemma}\label{lemma1}
Any successive transformation of the federation partition, i.e., ${\Im ^{(h)}} \to{\Im ^{(h+1)}},h=\{0,1,\cdots,H-1\}$, can improve the social welfare for miners in $\mathcal{M}_\tau$.
\end{lemma}

\begin{IEEEproof}
According to Definitions 6--7, for each switch operation $\Phi_{l,k}(m)$, it results in a strictly positive gain $\Omega_m(\Phi_{l,k})>0$ for the two involved pools $\Im_j$ and $\Im_k$, while the utilities of other pools in $\Im\backslash\{\Im_j,\Im_k\}$ remain unchanged.
Let $\sum^{(h)}$ be the set of permitted switch operations during the transformation ${\Im ^{(h)}} \to{\Im ^{(h+1)}}$. Therefore, we can obtain
\begin{align}\label{switchtransfain}
W\big(\Im^{(h+1)}\big)= W\big(\Im^{(h)}\big) + \sum_{\Phi_{l,k}(m) \in\sum^{(h)}} {\Omega_m(\Phi_{l,k})}.
\end{align}
The second term in the right side of (\ref{switchtransfain}) means the improved social welfare (i.e., the sum of added switch gains) during ${\Im ^{(h)}} \to{\Im ^{(h+1)}}$. As $\sum_{\Phi_{l,k}(m) \in\sum^{(h)}} {\Omega_m(\Phi_{l,k})}>0$, we have $W(\Im^{(h+1)})> W(\Im^{(h)})$. Lemma~\ref{lemma1} is proved.
\end{IEEEproof}

In the following, we first prove the convergence of federation partition transformations in Theorem~\ref{theorem2}. Then, we prove the Nash-stability, near-optimality, and low computational complexity in Theorems~\ref{theorem3}, \ref{theorem4}, \ref{theorem5}, respectively.

\begin{theorem}\label{theorem2}
Given an arbitrary initial pooled structure $\Im^{(0)}$, the proposed Algorithm~\ref{Algorithm2} can always converge to a final disjoint federation partition ${\Im}^*$ after a finite number of transformations.
\end{theorem}

\begin{IEEEproof}
According to the miner's switch strategy defined in Eq. (\ref{minerstrategy}), we can observe that a single switch operation either yields (i) a previously visited partition with a non-cooperative miner (i.e., a singleton) or (ii) an unvisited new partition.

\underline{Case (i).} When it comes to the partition structure $\Im^{(h)}$ where miner $m$ forms a singleton, the non-cooperative miner $m$ should either join a new federation or decide to remain non-cooperative in the next iteration $\Im^{(h+1)}$.
\begin{itemize}
  \item If miner $m$ chooses to join a new federation, an unvisited partition without this non-cooperative miner $m$ will be formed via the switch operation.
  \item If miner $m$ chooses to remain non-cooperative, any visited partition will not appear in the transformation of the current partition structure.
\end{itemize}

\underline{Case (ii).} When it comes to the partition structure $\Im^{(h)}$ where a single switch operation leads to an unvisited partition in the next iteration $\Im^{(h+1)}$, the maximum number of partitions among miners in $\mathcal{M}_\tau$ can be given by the well-known Bell number function, i.e.,
\begin{align}\label{eq:bell}
{b_{|\mathcal{M}_\tau|}} = \sum\nolimits_{m = 1}^{{|\mathcal{M}_\tau|}} {\left( \begin{array}{l}
{|\mathcal{M}_\tau|} \!-\! 1\\
~~~m
\end{array} \right)} {b_m},\forall m \in \mathcal{M}_\tau \ \&\ {b_0} = 1.
\end{align}

According to Eq. (\ref{eq:bell}), the number of transformations of federation partitions in formula (\ref{switchsequence}) is finite. Thereby, in all cases, the transformation sequence in formula (\ref{switchsequence}) will always terminate and converge to a final partition ${\Im}^*={\Im ^{(H)}}$, which is composed of a number of disjoint federations after finite $H$ iterations.
Theorem~\ref{theorem2} is proved.
\end{IEEEproof}

\emph{\textbf{Definition 9} (Nash-stability):} A federation partition $\Im = \{\Im_1,\cdots,\Im_J\}$ is Nash-stable if $\forall m \in \mathcal{M}_\tau, \forall \Im_j \in\Im, \forall \Im_{l} \in \Im \cup \{\emptyset\}, \Omega_m(\Phi_{j,l}) \leq 0$.

\begin{remark}
The Nash-stability implies that, under the Nash-stable partition, any single switch operation will not result in a strictly positive gain.
In other words, no miner has incentives to move from its current federation to another one or to deviate and act alone in the Nash-stable partition.
\end{remark}

\begin{theorem}\label{theorem3}
The final partition ${\Im}^*$ from Algorithm~\ref{Algorithm2} is Nash-stable.
\end{theorem}

\begin{IEEEproof}
The Nash-stability of partition ${\Im ^*}$ is proved by contradiction.
Suppose that the final partition ${\Im ^*}$ resulting from Algorithm~\ref{Algorithm2} is not Nash-stable. As a consequence, there exist a miner $m \in \Im_{j}$ and a federation $\Im_{j'} \in {\Im ^*}\cup \{\emptyset \}, j\ne j'$ such that $\Phi_{j,j'}(m){ \succeq }\Phi_{j,j}(m) $. Under this circumstance, miner $m$ will execute the switch operation $\Phi_{j,j'}(m)$, which contradicts with the fact that ${\Im ^*}$ is the final partition result of Algorithm~\ref{Algorithm2}. Theorem~\ref{theorem3} is proved.
\end{IEEEproof}

\begin{theorem}\label{theorem4}
The final partition outcome from Algorithm~\ref{Algorithm2} attains a near-optimal performance.
\end{theorem}

\begin{IEEEproof}
According to Lemma~\ref{lemma1}, the consecutive transformation of the federation partition, i.e., ${\Im ^{(h)}} \to{\Im ^{(h+1)}}$, can improve the overall social welfare. Moreover, according to Theorem~\ref{theorem2}, the overall system welfare derived by Algorithm~\ref{Algorithm2} is convergent after several iterations.
Besides, instead of allowing multiple miners to simultaneously transfer to a single pool, the switch operation to a single pool is executed in a greedy manner in our work, where only the miner with the largest positive switch gain is permitted.
Thereby, the partition result derived through Algorithm~\ref{Algorithm2} is near-optimal, which is also evaluated using simulations in comparison with the exhaustive optimal solution in Sect.~\ref{subsec:evalution2}.
\end{IEEEproof}

\begin{theorem}\label{theorem5}
The computational complexity of the proposed Algorithm~\ref{Algorithm2} is $\mathcal{O}( H\cdot J\cdot {\left| {{\mathcal{M}_\tau}} \right|})$, where $J=\left|{\Im}^*\right|$ is the number of federations in the Nash-stable partition ${\Im}^*$.
\end{theorem}

\begin{IEEEproof}
The complexity of the distributed pool formation algorithm in Algorithm~\ref{Algorithm2} mainly consists of two parts: (i) the first for-loop for each miner to decide its switch strategy (lines 5-8); and (ii) the second for-loop for each federation to decide its admission strategy (lines 9-15). For the first part, it has a complexity of $\mathcal{O}( H {\left| {{\mathcal{M}_\tau}} \right|}J)$. The second part has a complexity of $\mathcal{O}( H J{\left| \mathcal{C}_j^{\mathrm{miner}} \right|})$. As $\max\{ \left|\mathcal{C}_j^{\mathrm{miner}}\right|\} \le {\left| {{\mathcal{M}_\tau}} \right|}$, the overall computational complexity yields $\mathcal{O}( H J {\left| {{\mathcal{M}_\tau}} \right|})$.
Theorem~\ref{theorem5} is proved.
\end{IEEEproof}

\section{Performance Evaluation}\label{sec:SIMULATION}
In this section, the simulation setup is first introduced in Sect.~\ref{subsec:evalution0}, followed by the numerical results and discussions in Sects.~\ref{subsec:evalution1}$\sim$\ref{subsec:evalution3}.

\subsection{Simulation Setup}\label{subsec:evalution0}
We implement the prototype of PF-PoFL on a private blockchain network based on the open-source Go-Algorand project \cite{Go-Algorand} under both local and distributed settings. For the distributed setting (i.e., \textbf{setting 1}), the prototype is deployed across 20 Azure B8ms virtual machines (VMs), with 8 CPU cores and 32 GB of RAM. The VMs spread across five locations: West US, Canada East, UK South, Korea South, and North Europe. Each VM hosts 5 blockchain miners, and the total number of blockchain miners (i.e., $N$) is 100.
For the local setting (i.e., \textbf{setting 2}), the prototype is deployed and tested on a single computer with Intel Core i7-8700 CPU (3.6GHz) and 32GB RAM, where the number of blockchain miners is set as 100.

We use Go 1.17 to create and manage the private multi-node networks, as well as handle all networking and distributed systems aspects. We use the Turing-complete PyTeal \cite{PyTeal} to write the smart contract in Python 3.10 and use the \emph{compileProgram} method to produce the bytecode-based TEAL source code, which can be deployed on the blockchain. Besides, miners in a pool use PyTorch to perform federated mining with UDP to produce SGD updates during training, and our prototype can support any AI model which can be optimized via SGD.

\textbf{Dataset.} We consider two types of FL tasks trained on two typical datasets: one type is the handwritten digits recognition tasks on the MNIST dataset \cite{726791}; and another is the image recognition tasks on the CIFAR-10 dataset \cite{CIFAR10}.
MNIST consists of 60,000 training images and 10,000 test images with size of $28\times28$ in 10 classes.
CIFAR-10 contains 60,000 $32\times32$ images evenly divided in 10 classes, with 50,000 training images and 10,000 test samples.
For the dataset partition among miners, the non-IID setting is adopted. Specifically, each miner is assigned with $z$ classes of 600 training samples for MNIST (resp. 500 training samples for CIFAR-10) and the training samples of different miners can be repeated, where the integer $z$ is randomly selected from $\{1,2,\cdots,10\}$.

\textbf{Model.} For federated mining within each pool in PF-PoFL, the 3-layer CNN model is adopted for MNIST, while the ResNet18 model is employed for CIFAR-10.
The total number of communication rounds is set as $\Psi_{j,\tau}^{\mathrm{global}}=200$, and the sample ratio is set as $\lambda_s = 0.35$.
For the learning algorithm, the mini-batch SGD with local batch size equal to 50 is adopted for all miners, where the initial learning rate $\eta$ is set as 0.01 and declines to 0.001 after 120-th communication round. Besides, the local epoch of miners is set as 1 under the federated mining strategy, while the maximum local epoch $\Psi_{\max,\tau}^{\mathrm{local}}$ of miners is set as 1000 under the solo mining strategy.

Table~\ref{dataset} summarizes the datasets and models used in PF-PoFL. For the UDP model, the privacy parameters $(\sigma,\epsilon)$ are selected from $\{(3,5.52), (14,1.06), (24,0.61)\}$. Besides, we set $\delta=1\times 10^{-6}$. For the federation formation game model, we set $\gamma_d=30$, $\gamma_s=23$, $\lambda_c=0.01$. In addition, we set miner's initial credit value as $cre_n^0=1$ and the credit updates as $\chi_1=2$, $\chi_2=4$, and $\chi_3=1$.

\begin{table}[!t]
\caption{Used datasets and models in PF-PoFL}\label{dataset}\vspace{-0.7cm}
\begin{center}
\begin{tabular}{c|cccc}\hline
Dataset  & \#Training samples & \#Test samples & \#Classes & Model    \\ \hline
MNIST    & 60,000             & 10,000         & 10        & CNN      \\
CIFAR-10 & 50,000             & 10,000         & 10        & ResNet18 \\ \hline
\end{tabular}
\end{center} \vspace{-1mm}
\end{table}

\textbf{Comparing method.} We evaluate the performance of the PF-PoFL in comparison with the PoFL scheme \cite{9347812}. In PoFL \cite{9347812}, its working relies on a central FL platform, and the fixed mining pool structure is employed for federated mining. PoFL ensures privacy preservation and prevents model plagiarism for trained models using homomorphic encryption (HE)-based label prediction and secure two-party computation (2PC)-based label comparison. As PoFL does not specify the detailed agreement protocol to acquire consensus, to be objective and fair, we implement the conventional Algorand BA protocol \cite{Gilad2017Algorand} for miners in PoFL to reach consensus on a new block. Here, the number of miners is set as $N=100$ and the number of malicious or Byzantine miners is set as $f=20$.

In the following, we first evaluate the blockchain throughput, block latency, and system overhead of our PF-PoFL in Figs.~\ref{fig:bc1}--\ref{fig:bc2} and Tables~\ref{timeCostMRC}--\ref{timeCost}.
Next, we analyze the impacts of non-IID, number of miners, and privacy parameters on the federated mining process with UDP in Figs.~\ref{fig:1}--\ref{fig:3-2}. Finally, we analyze the efficiency and stability of the federation formation game approach for optimized pool structure in Figs.~\ref{fig:4}--\ref{fig:5}.

\subsection{System Throughput and Latency}\label{subsec:evalution1}
\begin{figure}[!t]\centering\setlength{\abovecaptionskip}{-0.01cm}
\includegraphics[width=9cm]{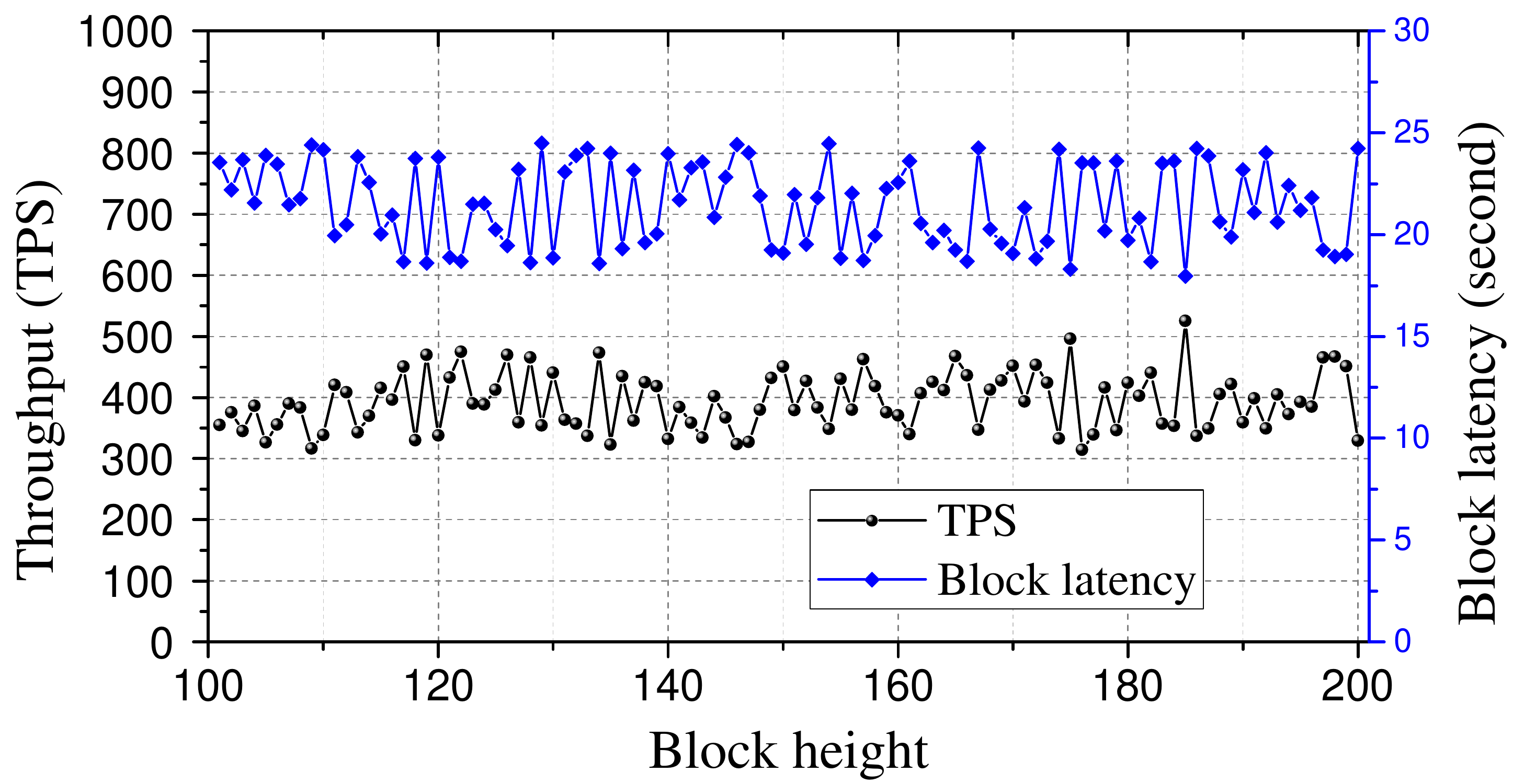}
\caption{Throughput and block latency in different block heights in PF-PoFL (setting 1).}
\label{fig:bc1}\vspace{-1mm}
\end{figure}

Fig.~\ref{fig:bc1} shows the throughput (measured by transactions per second (TPS)) and the block latency (i.e., the consensus time for a block from being pending to be confirmed) of the blockchain system in the distributed setting.
In the blockchain, the shorter the consensus time, the better the consensus efficiency. Here, the consensus time is mainly determined by the delay in model downloading, model ranking, model verification, block propagation, and voting-based agreement process. In the simulation, the ratio of malicious or Byzantine miners is set to be 20\%.
From Fig.~\ref{fig:bc1}, it can be seen that, when the block height grows from 100 to 200, both the TPS and the block latency of our proposed PF-PoFL consensus mechanism remain relatively stable. Specifically, the throughput mainly varies between [300, 500] TPS, while the block latency changes within [18, 25] seconds, which indicates a relatively stable block packaging frequency of our PF-PoFL consensus mechanism. It can be explained as follows.

In the traditional PoW consensus protocol, all miners compete to solve the same PoW puzzle and the fastest node to solve it earns the block rewards. Distinguished from the PoW protocol, in our PF-PoFL consensus protocol, different mining pools can concurrently train on different FL tasks that are in the training phase; meanwhile, in the current block height, the validators prefer to reach consensus on the most urgent and profitable task from all tasks that are in the testing phase. For example, for a FL task $\tau$ with a high difficulty level, in its training phase, validators will carry out consensus on other tasks that are in the testing phase; until the FL task $\tau$ reaches its testing phase, validators will perform consensus on it if the task $\tau$ is most urgent and profitable in current block height. As such, for FL tasks with different difficulty levels, the relatively stable throughput of the blockchain system can be obtained.

\begin{figure}[!t]\centering\setlength{\abovecaptionskip}{-0.01cm}
\includegraphics[width=8.6cm]{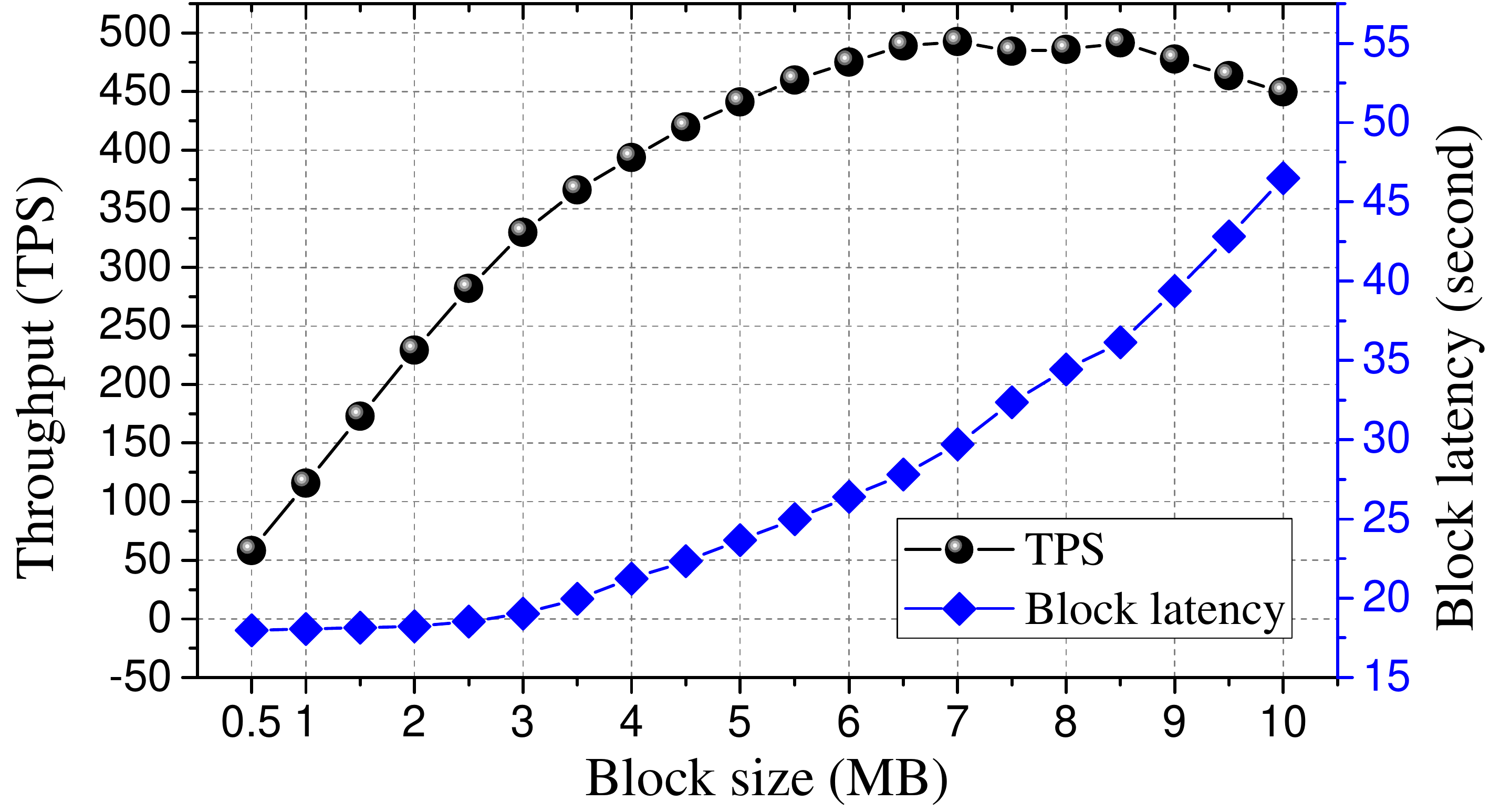}
\caption{Throughput and block latency vs. block size in PF-PoFL (setting 1).}
\label{fig:bc2}\vspace{-2mm}
\end{figure}

Fig.~\ref{fig:bc2} shows the throughput and the block latency of PF-PoFL in the distributed setting when the block size increases from 0.5 MB to 10 MB. It can be observed that our PF-PoFL system can efficiently handle hundreds of transactions per second and the block latency is below 47 seconds. As seen in Fig.~\ref{fig:bc2}, with the increase of block size, the block latency keeps increasing while the blockchain throughput first increases then decreases when the block size is over 8.5 MB. It can be explained as follows. On one hand, the higher block size indicates that more transactions are included within a block, resulting in a higher latency in transaction processing and block propagation. On the other hand, the relatively higher block size means that more transactions to be processed in a new block, which brings a higher TPS. However, the TPS cannot grow with the block size all the time. The reason is when the block size is too large (i.e., over 8.5 MB), the time to verify and broadcast the block can be extended, so that the distributed blockchain cannot be synchronized in time.

\begin{table}[!t]
\caption{Time cost of executing model ranking contract in PF-PoFL (setting 1)}\label{timeCostMRC}\vspace{-0.2cm}
\begin{tabular}{c||ccc}
\hline
Operation          & \multicolumn{1}{c|}{\begin{tabular}[c]{@{}c@{}}Time for the \\ fastest validator\end{tabular}} & \multicolumn{1}{c|}{\begin{tabular}[c]{@{}c@{}}Time for the \\ slowest validator\end{tabular}} & \begin{tabular}[c]{@{}c@{}}Ave. time for \\ validators\end{tabular} \\ \hline
{Running}            & \multicolumn{1}{c|}{3.215 s}                                                                   & \multicolumn{1}{c|}{6.246 s}                                                                   & 4.952 s                                                             \\ \hline
\begin{tabular}[c]{@{}c@{}}{Reaching}\\{agreement}\end{tabular} & \multicolumn{3}{c}{16.583 s}                                                                                                                                                                                                                                         \\ \hline
\textbf{Overall latency}    & \multicolumn{3}{c}{$\thickapprox$\,21.1 s}                                                                                                                                                                                                                                           \\ \hline
\end{tabular} \vspace{-0.12cm}
\end{table}

\begin{table*}[htbp]
\caption{Computation cost of model operations for pools and validators under PoFL and our PF-PoFL}\label{timeCost}\vspace{-0.33cm}
\begin{center} 
\begin{tabular}{c|c||c||cccc}
\hline
\multirow{2}{*}{Setting}            & \multirow{2}{*}{Scheme} & \textbf{Pool}    & \multicolumn{4}{c}{\textbf{Validator}}                                                                                                       \\ \cline{3-7}
                                    &                         & Model evaluation & \multicolumn{1}{c|}{Downloading} & \multicolumn{1}{c|}{Model verification} & \multicolumn{1}{c||}{Model ranking} & \textbf{Total consensus delay} \\ \hline
\multirow{2}{*}{\textbf{Setting 1}} & \textbf{PoFL \cite{9347812}}           & $1.7 \times 10^6$ s     & \multicolumn{1}{c|}{Negligible}  & \multicolumn{1}{c|}{1.675 s}            & \multicolumn{1}{c||}{59 us}         & 12.2 s              \\ \cline{2-7}
                                    & \textbf{Ours}           & Negligible       & \multicolumn{1}{c|}{Negligible}  & \multicolumn{1}{c|}{3.241 s}           & \multicolumn{1}{c||}{184 us}        & 9.4 s              \\ \hline
\multirow{2}{*}{\textbf{Setting 2}} & \textbf{PoFL \cite{9347812}}           & $2.3 \times 10^6$ s     & \multicolumn{1}{c|}{1.21 s}      & \multicolumn{1}{c|}{1.296 s}            & \multicolumn{1}{c||}{51 us}         & 39.8 s            \\ \cline{2-7}
                                    & \textbf{Ours}           & Negligible       & \multicolumn{1}{c|}{1.88 s}      & \multicolumn{1}{c|}{2.927 s}            & \multicolumn{1}{c||}{172 us}        & 21.3 s             \\ \hline
\end{tabular}
\end{center} \vspace{-0.3cm}
\end{table*}

Table~\ref{timeCostMRC} shows the time cost of executing model ranking contract in PF-PoFL under the distributed setting. The execution time of model ranking contract mainly consists of two parts: the contract running time (to produce local model ranking result) and the time to reach agreement on the final model ranking result (to be committed on blockchain). Note that in Algorithm 1, each validator can process multiple FL model transactions (including model downloading and evaluation) in parallel to reduce the contract running time. As seen in Table~\ref{timeCostMRC}, the average contract running time of validators is 4.952 seconds and the time to reach agreement is 16.583 seconds, leading to a overall latency of about 21.1 seconds.

Table~\ref{timeCost} compares the time cost of model operations for pools and validators in PoFL and PF-PoFL under both local and distributed settings.
As seen in Table~\ref{timeCost}, PoFL involves huge computation cost in model evaluation process (i.e., calculating model performance), especially for pools due to the expensive HE operations.
Compared with the PoFL scheme, our PF-PoFL adds the Gaussian noise according to Eqs. (\ref{eq:scaledupdate}) and (\ref{eq:globalagg}) to ensure UDP for miners, and the time cost for pools in this process can be negligible.
For model ranking and verification operations performed by validators, PoFL only needs to determine the best model. It indicates that if the model with the best performance is verified as true, there is no need to verify other models.
In our PF-PoFL, each validator needs to test and sort all relevant models to determine the financial rewards and credit incentives in the block rewarding contract, causing a longer running time (e.g., 172 microseconds for model ranking and 2.927 seconds for model verification in setting 2) than that in the PoFL (e.g., 51 microseconds for model ranking and  1.296 seconds for model verification in setting 2).
Besides, Table~\ref{timeCost} shows that the total consensus time for reaching agreement among validators on a new block is about 9.4 seconds in our PF-PoFL in setting 1 (resp. 21.3 seconds in setting 2), which is faster than that in the PoFL, i.e., 12.2 seconds in setting 1 (resp. 39.8 seconds in setting 2). The reasons are as follows. On one hand, the block proposal in PoFL involves huge storage space for encrypted model parameters in HE and 2PC, resulting in a higher block propagation latency. On the other hand, our credit-based Algorand BA protocol can employ more honest miners in the validator committee and incentivize miners to act legitimately and actively, thereby reducing the voting stages and the delay in reaching an unambiguous agreement.

\subsection{Federated Mining with UDP}\label{subsec:evalution2}
\begin{figure}[t]\centering\setlength{\abovecaptionskip}{-0.02cm}
\includegraphics[height=6cm]{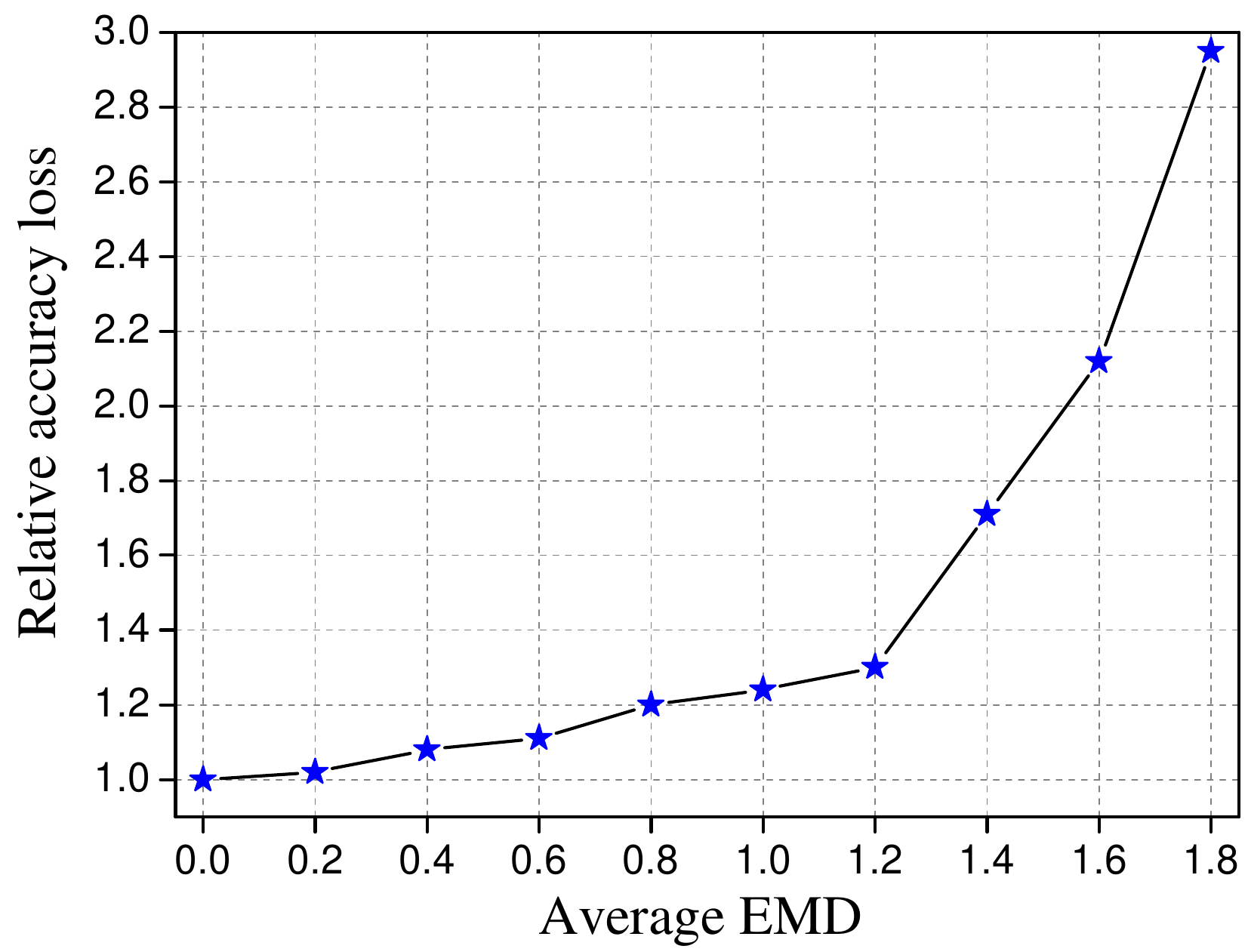}
\caption{Relative accuracy loss $\mathcal{E}(\overline{\Xi}_j)$ vs. average EMD $\overline{\Xi}_j$ (CIFAR-10).}
\label{fig:1}\vspace{-2mm}
\end{figure}

In this subsection, we evaluate the performance of the federated mining process with UDP on MNIST and CIFAR-10, in comparison with the baseline scheme, by changing the number of miners and values of privacy parameters. In the baseline scheme, the curator in a pool directly delivers the raw global update (instead of the perturbed version) to all miners.
Fig.~\ref{fig:1} illustrates the relationship between average EMD $\overline{\Xi}_j$ and relative accuracy loss $\mathcal{E}(\overline{\Xi}_j)$ in Eq. (\ref{accuracyloss}) on CIFAR-10 dataset under non-IID settings. As seen in Fig.~\ref{fig:1}, the relative accuracy loss first grows slowly and tends to increase fast when $\overline{\Xi}_j>1.2$. Besides, when $\overline{\Xi}_j=0$, it corresponds to the IID case and we have $\mathcal{E}(\overline{\Xi}_j)=1$. According to Eq. (\ref{nonIID}), a higher EMD indicates a higher non-IID degree, and thereby a larger drop on test accuracy.

\begin{figure}[t]\centering\setlength{\abovecaptionskip}{-0.01cm}
\includegraphics[height=6cm]{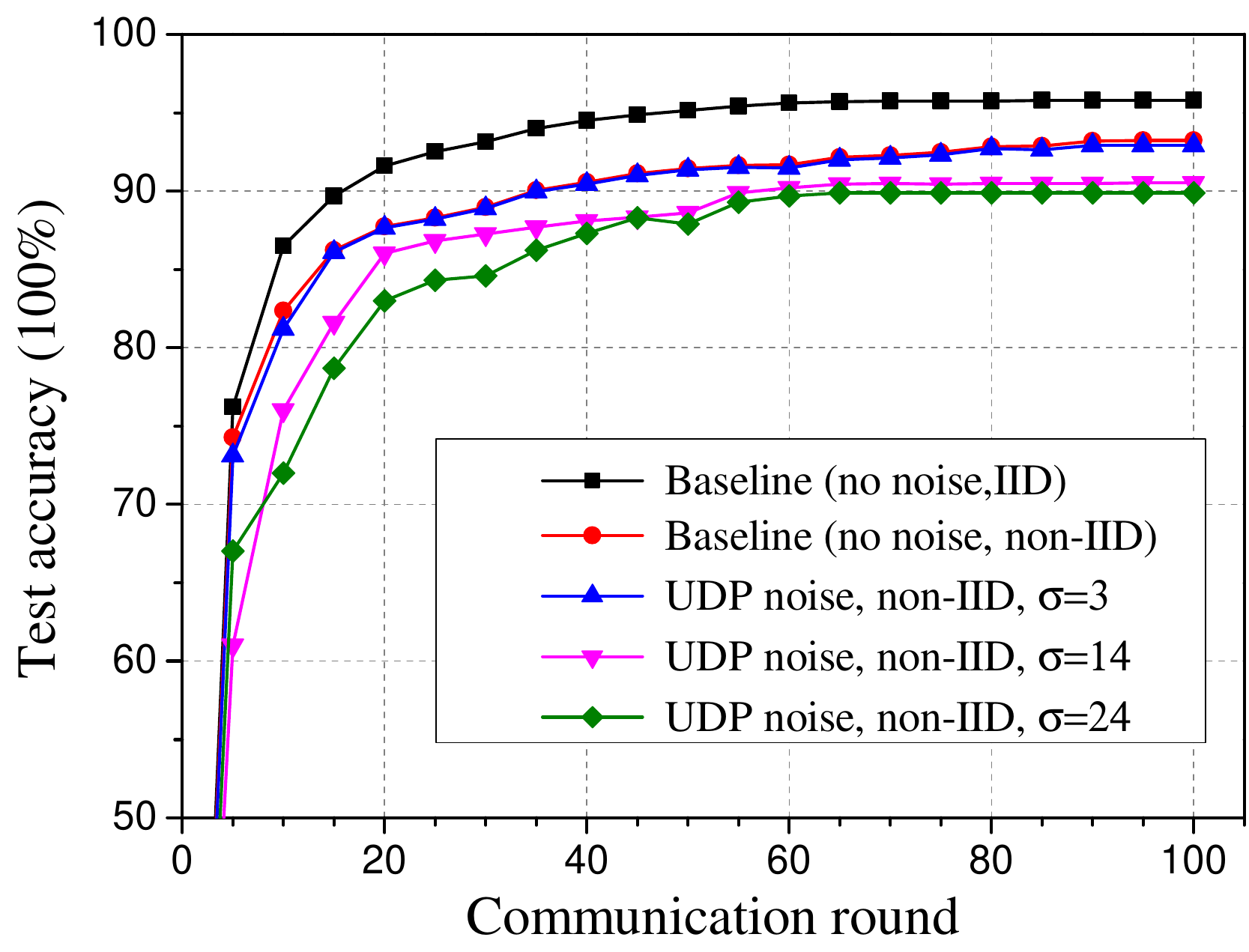}
\caption{Test accuracy of federated mining vs. noise parameter $\sigma$ and data distribution, under different communication rounds (MNIST, $N=100$).}
\label{fig:2-1}\vspace{-0mm}
\end{figure}

\begin{figure}[t]\centering\setlength{\abovecaptionskip}{-0.01cm}
\includegraphics[height=6cm]{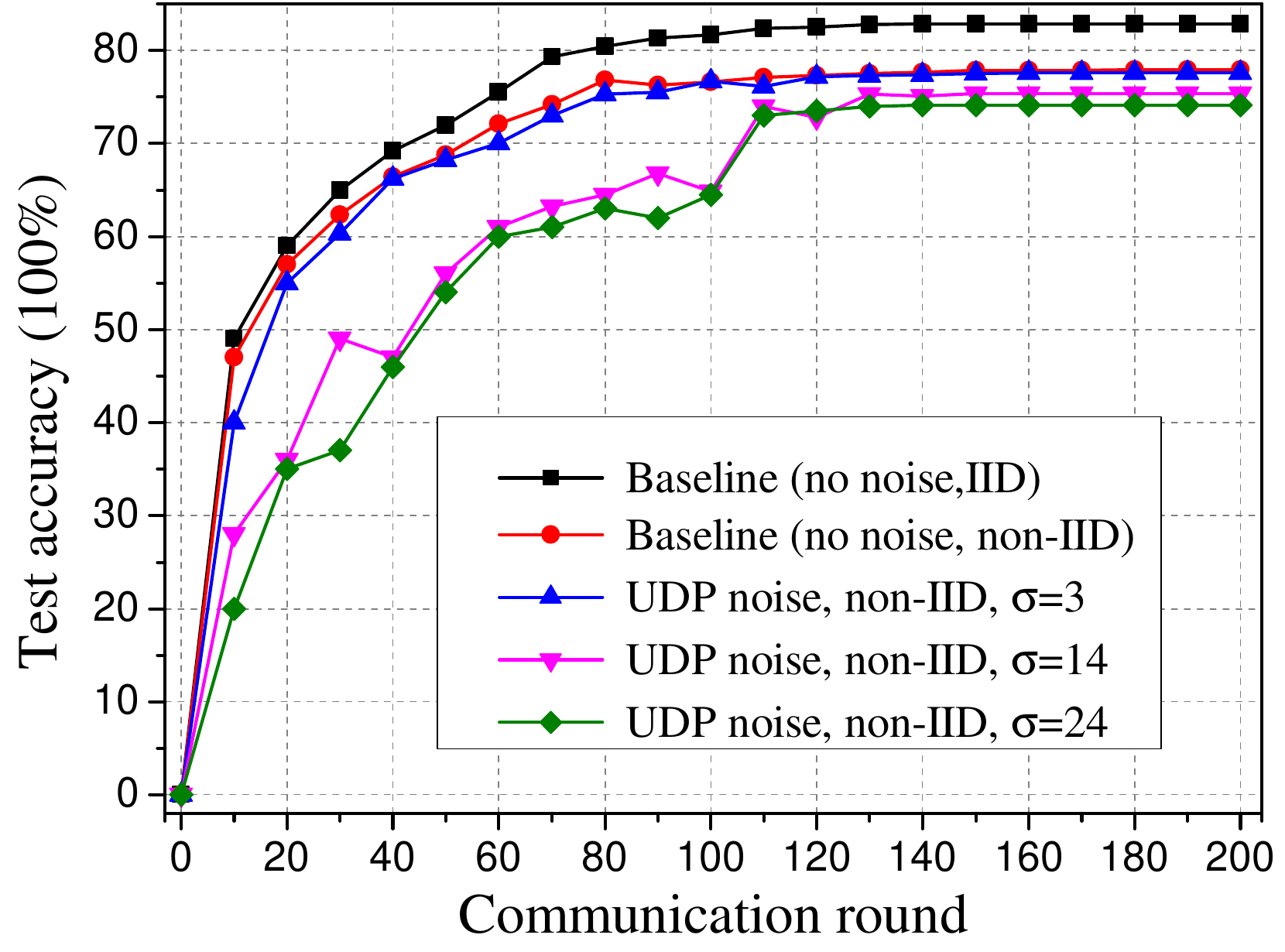}
\caption{Test accuracy of federated mining vs. noise parameter $\sigma$ and data distribution, under different communication rounds (CIFAR-10, $N=100$).}
\label{fig:2-2}\vspace{-2mm}
\end{figure}

Figs.~\ref{fig:2-1} and \ref{fig:2-2} show the evolution of test accuracy on MNIST and CIFAR-10 under different privacy parameters in our UDP noise-adding mechanism, respectively, compared with baseline schemes. In this simulation, the number of participating miners is fixed as 100.
As seen in Figs.~\ref{fig:2-1} and \ref{fig:2-2}, the test accuracy of the trained model via federated mining with $\sigma=3$ UDP noise is very close to the baseline under non-IID, while the accuracies of the trained model with $\sigma=14$ and $\sigma=24$ are decreased to a certain degree. It is because a higher $\sigma$ implies a higher privacy preservation level, causing larger perturbations to be added to the global model update in Eq. (\ref{eq:globalagg}) which eventually deteriorates the model accuracy. Moreover, compared with the IID case, the existence of non-IID results in an accuracy drop to the baseline.

\begin{figure}[t]\centering\setlength{\abovecaptionskip}{-0.01cm}
\includegraphics[height=6cm]{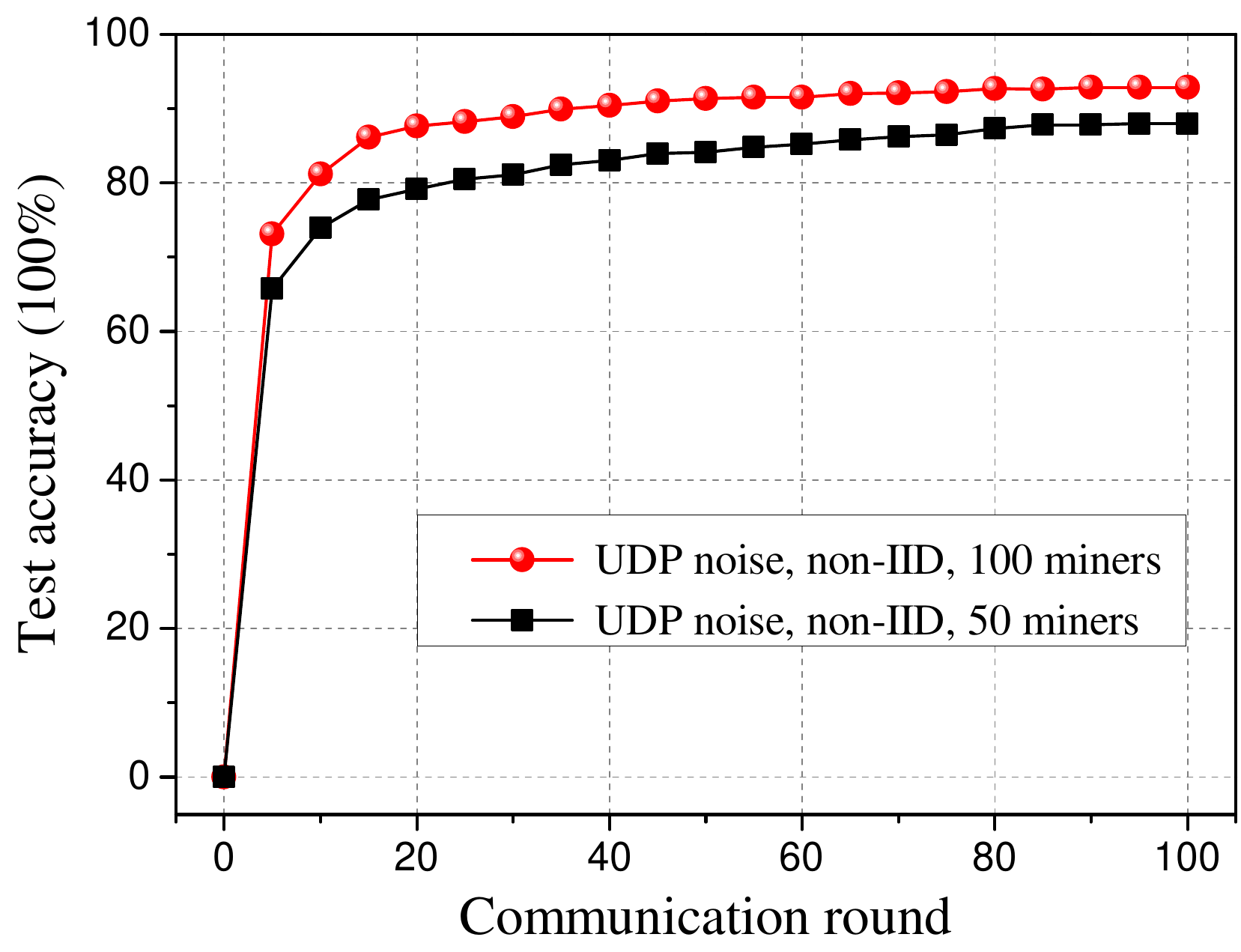}
\caption{Test accuracy of federated mining vs. number of participating miners ${|\Im_j|}$, under different communication rounds (MNIST, $\sigma=3$).}
\label{fig:3-1}\vspace{-1.1mm}
\end{figure}

\begin{figure}[t]\centering\setlength{\abovecaptionskip}{-0.01cm}
\includegraphics[height=6cm]{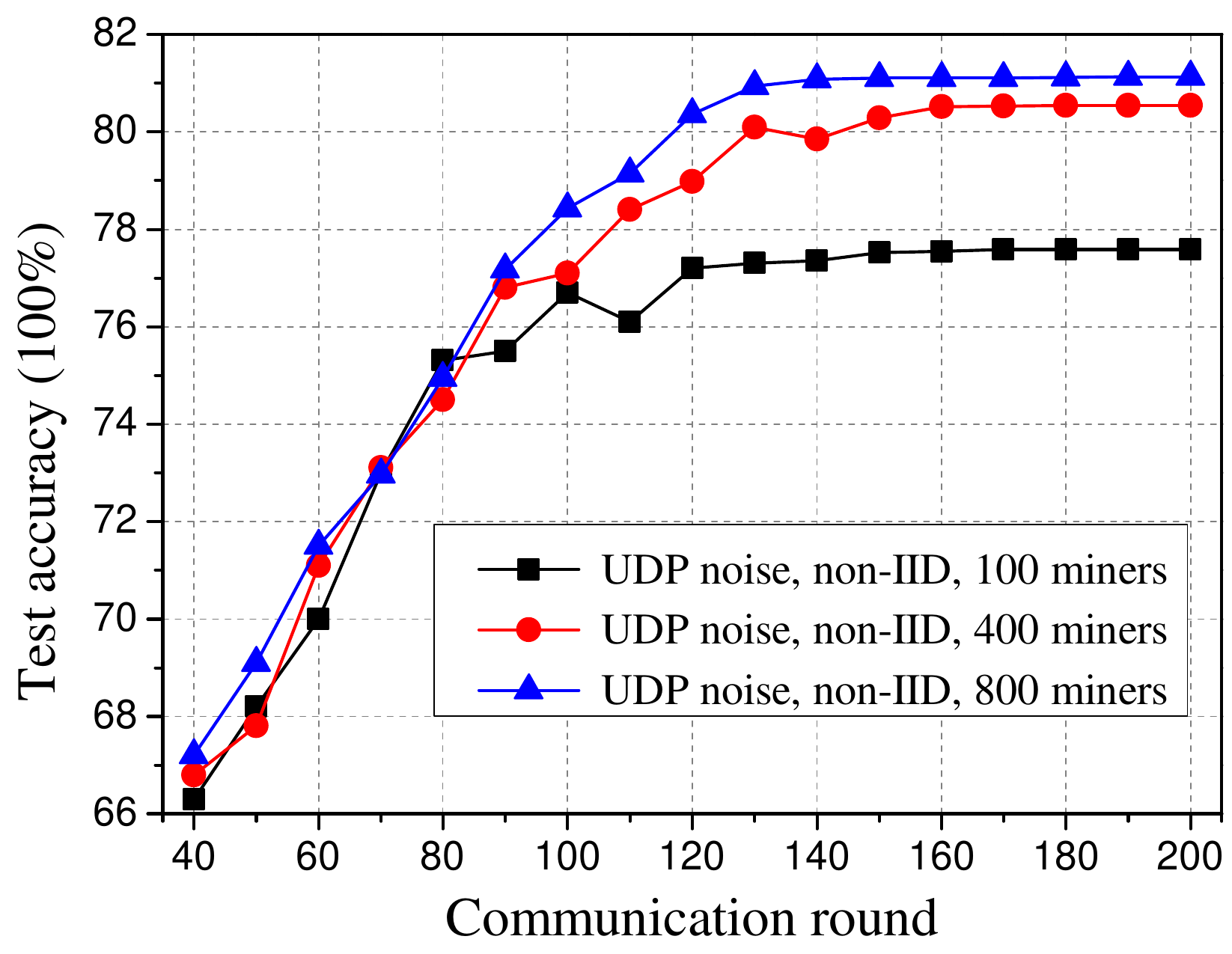}
\caption{Test accuracy of federated mining vs. number of participating miners ${|\Im_j|}$, under different communication rounds (CIFAR-10, $\sigma=3$).}
\label{fig:3-2}\vspace{-2.mm}
\end{figure}

Figs.~\ref{fig:3-1} and \ref{fig:3-2} show the evolution of test accuracy on MNIST and CIFAR-10 under different pool sizes (i.e., number of miners in a pool), respectively. In this simulation, we set $\sigma=3$. From Figs.~\ref{fig:3-1} and \ref{fig:3-2}, we can observe that with a larger pool size, the performance of the trained model is improving in terms of both accuracy and stability. The reason is that given the fixed sample ratio, when the number of participating miners in a pool increases, the impact of the added UDP noise to the global update in Eq. (\ref{eq:globalagg}) can be reduced, resulting in the higher accuracy and stability in training FL models.

\subsection{Federation Formation Game}\label{subsec:evalution3}
In this subsection, we evaluate the performance of the proposed federation formation game in comparison with the exhaustive optimal scheme, the non-cooperative scheme, and the PoFL scheme \cite{9347812}. In the exhaustive optimal scheme, the optimal pool structure among miners is derived via the exhaustive searching method in a centralized manner. In the non-cooperative scheme, each miner behaves uncooperatively and applies the solo-mining strategy in completing FL tasks. In PoFL scheme \cite{9347812}, all miners utilize the off-the-shelf fixed pooled-mining structure in PoW-based blockchains for federated mining. Here, the total number of miners is set as $N=100$, where miners perform federated mining on CIFAR-10 using the ResNet18 model under the setting 1.

\begin{figure}[t]\centering\setlength{\abovecaptionskip}{-0.01cm}
\includegraphics[height=6cm]{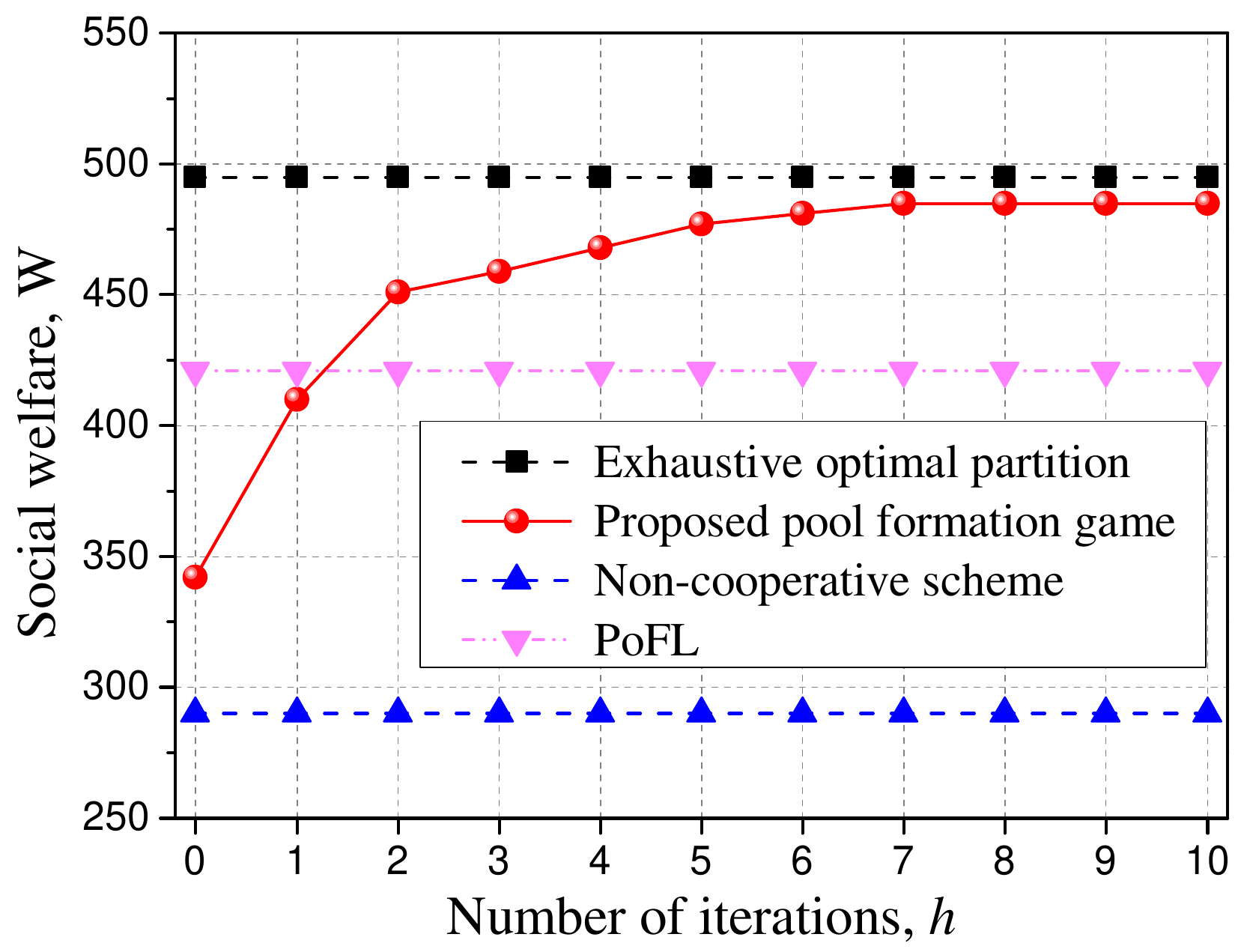}
\caption{Evolution of social welfare $W(\Im)$ of miner partition over time in four schemes ($N=100$).}
\label{fig:4}\vspace{-1.1mm}
\end{figure}

\begin{figure}[t]\centering\setlength{\abovecaptionskip}{-0.01cm}
~~\includegraphics[height=6cm]{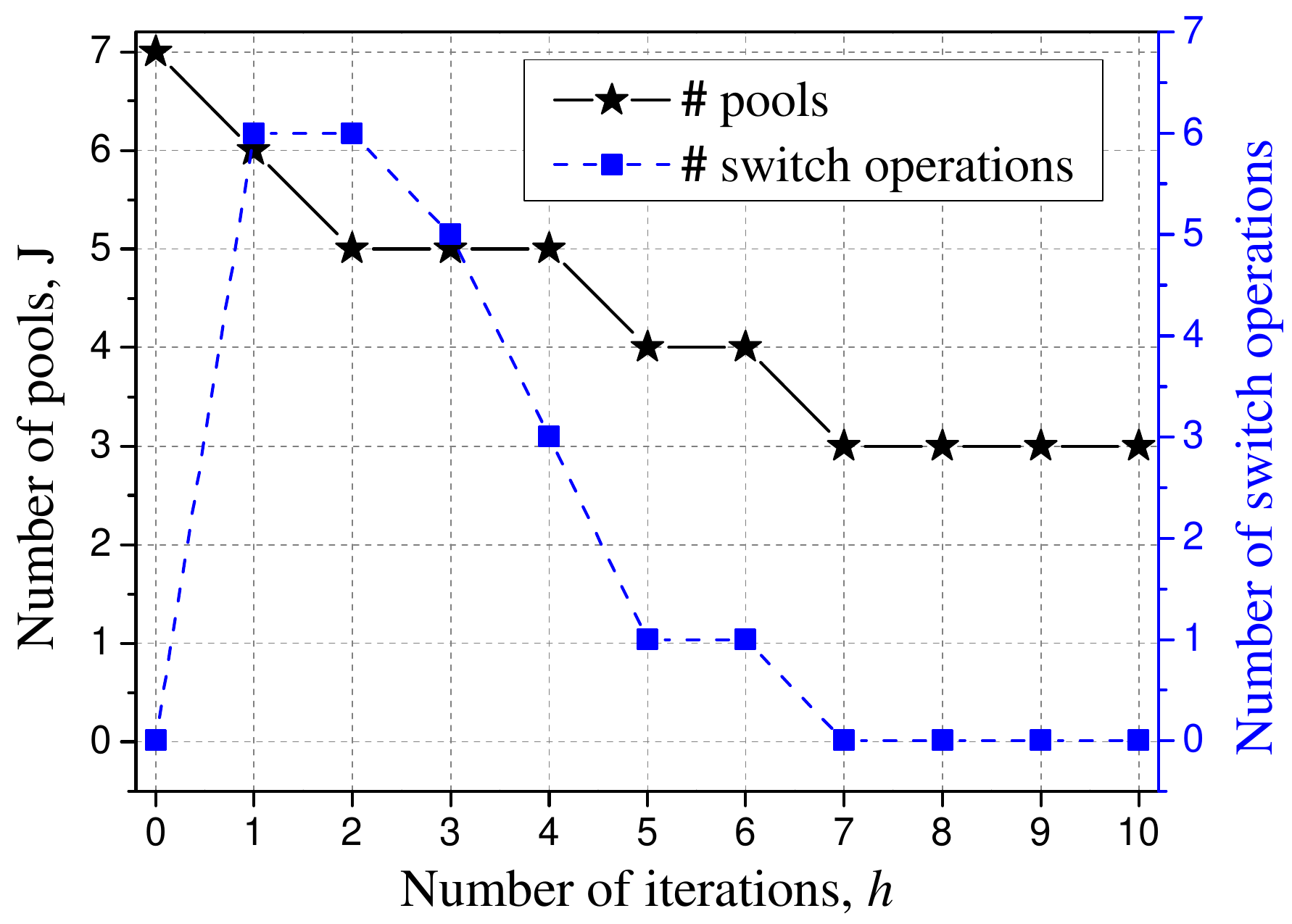}
\caption{Evolution of partition structure over time in terms of number of pools and number of switch operations ($N=100$).}
\label{fig:5}\vspace{-2mm}
\end{figure}

Fig.~\ref{fig:4} compares the social welfare $W$ defined in formula (\ref{socialwelfare}) in four schemes.
As shown in Fig.~\ref{fig:4}, when the number of iterations grows, the social welfare of miner partition in the proposed FFG-TU game gradually converges to a stable value after around 7 iterations. Besides, the proposed FFG-TU game can yield a near-optimal performance, as its performance gap to the exhaustive optimal solution does not exceed 2.1\% when $h\geq 7$.
In addition, as seen in Fig.~\ref{fig:4}, the stable social welfare in our FFG-TU game is greater than that in the non-cooperative scheme and the PoFL scheme, and it brings up to about 67\% improvement to the non-cooperative scheme and about 17.6\% improvement to the PoFL scheme when $h=10$. The reason is that in the non-cooperative scheme, all miners work alone and each of them forms a singleton in conducting FL tasks, resulting in the lowest social welfare. Besides, as different miners generally have distinct training samples and non-IID degrees under heterogenous FL tasks, the fixed mining pool structure in the PoFL scheme cannot adapt to the varying federated mining environment, resulting in relatively lower social welfare.

Fig.~\ref{fig:5} further inspects the evolution of partition structure among miners over time in terms of the number of pools and the number of switch operations. As seen in Figs.~\ref{fig:4} and \ref{fig:5}, the switch operations at each iteration can improve the social welfare for miners, which accords with Lemma~\ref{lemma1}. Moreover, as observed in Fig.~\ref{fig:5}, the proposed pool formation algorithm converges to a final disjoint partition structure after 7 iterations and achieves Nash-stability, which accords with Theorems~\ref{theorem2} and \ref{theorem3}. Besides, Fig.~\ref{fig:5} shows that, the number of switch operations $|\sum^{(h)}|$ at each iteration $h$ satisfies $|\sum^{(h)}|\leq |\Im^{(h)}| +1$, where $|\Im^{(h)}|$ is the number of pools in current partition $\Im^{(h)}$. The reason is that, according to the switch rule and admission rule in Definitions~7 and 8, each miner can either switch to another pool or form a singleton if it decides to leave the current pool, while each pool only admits the optimal miner greedily at each iteration.

According to the aforementioned results, our PF-PoFL can effectively recycle energy for efficient federated mining and attain stable throughput of the blockchain system, low block latency in reaching consensus, desirable model performance with UDP guarantees, and optimized federated structure with high social welfare.

\section{Conclusion}\label{sec:CONSLUSION}
In this paper, we have proposed a novel energy-recycling consensus mechanism named PF-PoFL, where the wasted energy in solving cryptographic puzzles in PoW is reinvested to FL.
To implement PF-PoFL in a fully decentralized manner, we utilize the blockchain to host the unfinished tasks, perform model ranking, and enforce financial settlement by devising a novel block structure, a model ranking contract, and a block rewarding contract. PF-PoFL also incorporates credit-based incentives to motivate miners' honest participation and improve consensus efficiency. Furthermore, a user-level privacy-preserving model training algorithm is designed to offer rigorous privacy protection for miners in each pool. In addition, based on the federation formation game, we present an optimized pool formulation algorithm, where miners with diverse characteristics (i.e., training samples, non-IID degree, and network delay) in heterogenous FL tasks can self-organize into a disjoint Nash-stable partition.
Simulation results have validated the efficiency and effectiveness of the proposed PF-PoFL mechanism.
For the future work, we will further investigate the optimized PF-PoFL consensus mechanism with optimal FL task rewarding, task difficulty adjustment function, and non-reusability guarantees.

\bibliographystyle{IEEETran}
\bibliography{Ref_PoFL}

\begin{IEEEbiography}[{\includegraphics[width=1in,height=1.25in,clip,keepaspectratio]{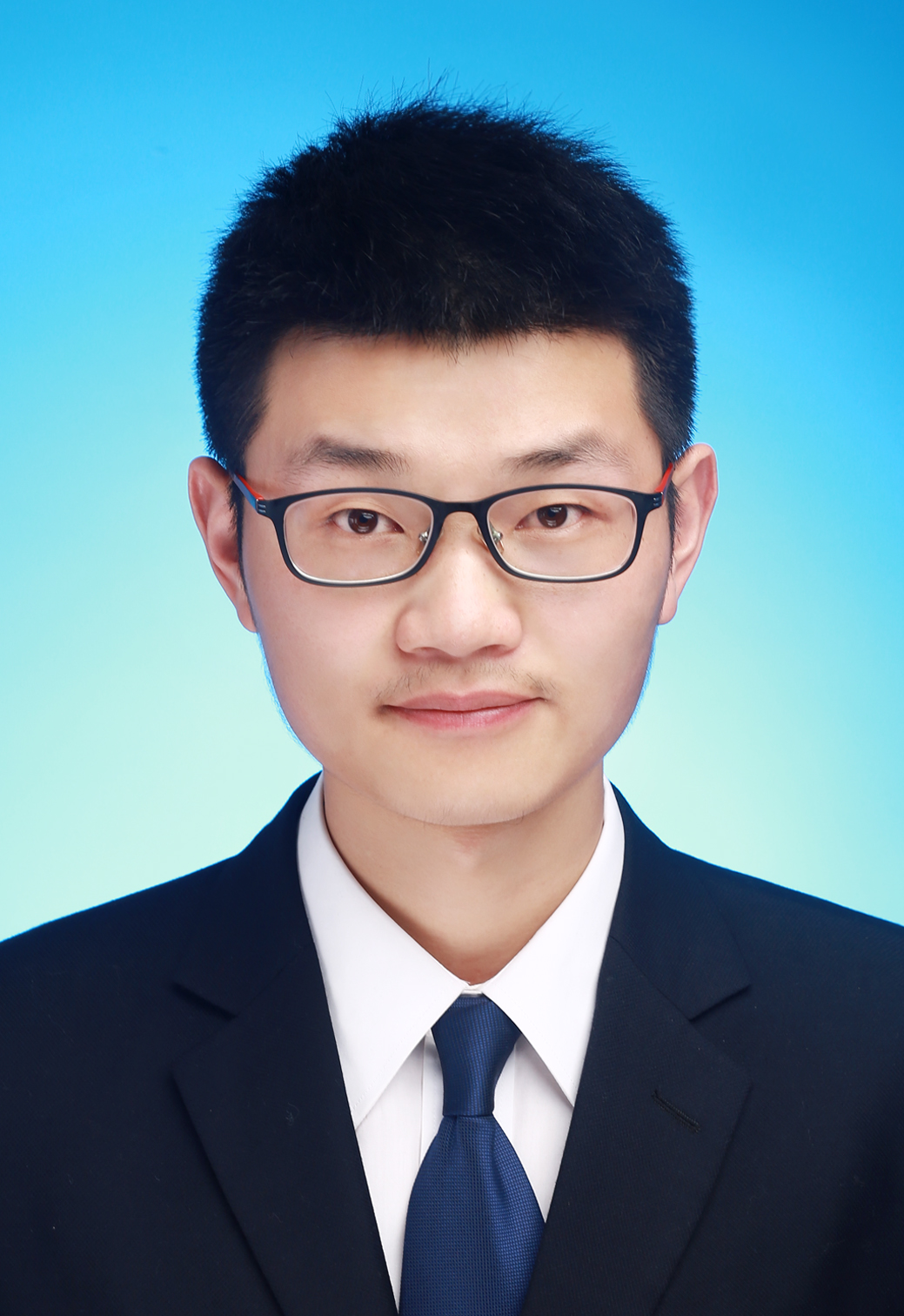}}]{Yuntao Wang}
received the Ph.D degree in Cyber Science and Engineering from Xi'an Jiaotong University, Xi'an, China, in 2022. His research interests include security and privacy protection in wireless networks, vehicular networks, and UAV networks.
\end{IEEEbiography}\vspace{-5mm}

\begin{IEEEbiography}[{\includegraphics[width=1in,height=1.25in,clip,keepaspectratio]{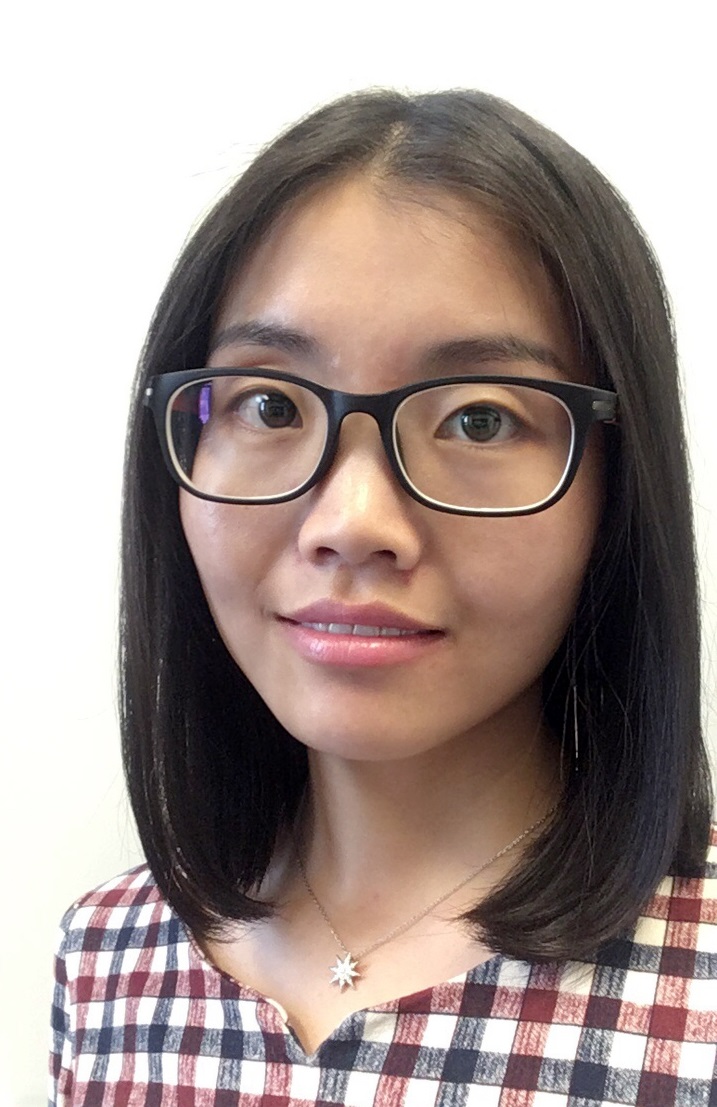}}]{Haixia Peng}
received her Ph.D. degrees in Computer Science and Electrical \& Computer Engineering from Northeastern University (Shenyang, China) in 2017 and the University of Waterloo (Waterloo, Canada) in 2021, respectively. 
She is currently an Associate Professor with the School of Information and Communications Engineering, Xi'an Jiaotong University, China. Her research interests include satellite-terrestrial vehicular networks, multi-access edge computing, resource management, artificial intelligence, and reinforcement learning. She serves/served as a TPC member in IEEE VTC-fall 2016\&2017, IEEE ICCEREC 2018, IEEE GlobeCom 2016-2022, and IEEE ICC 2017-2022 conferences and serves as an Associate Editor for the {\scshape Peer-to-Peer Networking and Applications}.
\end{IEEEbiography}\vspace{-5mm}

\begin{IEEEbiography}[{\includegraphics[width=1in,height=1.25in,clip,keepaspectratio]{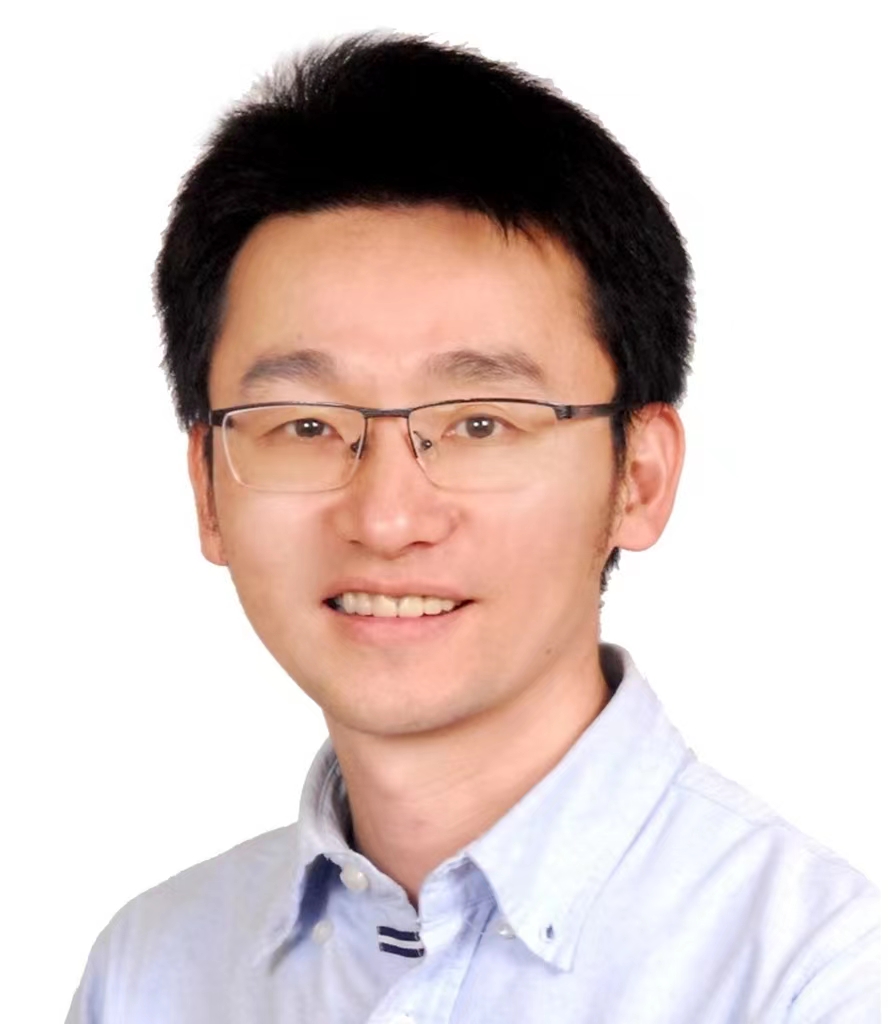}}]{Zhou Su}
has published technical papers, including top journals and top conferences, such as {\scshape IEEE Journal On Selected Areas In Communications}, {\scshape IEEE Transactions On Information Forensics And Security}, {\scshape IEEE Transactions On Dependable And Secure Computing}, {\scshape IEEE Transactions On Mobile Computing}, {\scshape IEEE/ACM Transactions On Networking}, and {\scshape INFOCOM}. His research interests include multimedia communication, wireless communication, and network traffic. Dr. Su received the Best Paper Award of International Conference IEEE ICC2020, IEEE BigdataSE2019, and IEEE CyberSciTech2017. He is an Associate Editor of {\scshape IEEE Internet Of Things Journal}, {\scshape IEEE Open Journal Of Computer Society}, and {\scshape IET Communications}.
\end{IEEEbiography}\vspace{-5mm}

\begin{IEEEbiography}[{\includegraphics[width=1in,height=1.25in,clip,keepaspectratio]{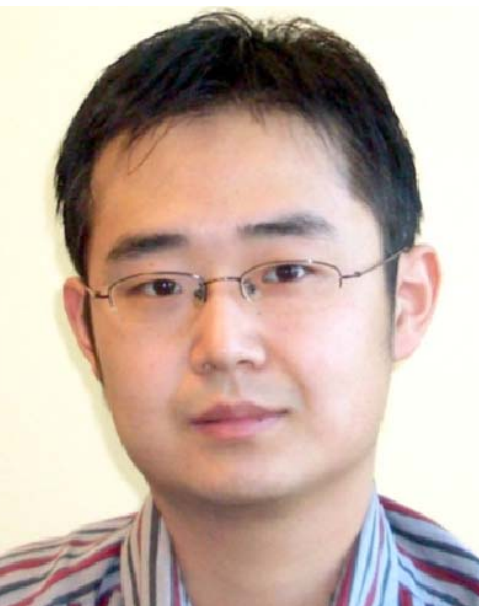}}]{Tom H. Luan}
received the Ph.D. degree from the University of Waterloo, Canada, in 2012. He is currently a Professor with the School of Cyber Science and Engineering, Xi'an Jiaotong University, China. His research mainly focuses on content distribution and media streaming in vehicular ad hoc networks and peer-to-peer networking and the protocol design and performance evaluation of wireless cloud computing and edge computing. He served as a TPC Member for IEEE Globecom, ICC, and PIMRC.
\end{IEEEbiography}\vspace{-5mm}

\begin{IEEEbiography}[{\includegraphics[width=1in,height=1.25in,clip,keepaspectratio]{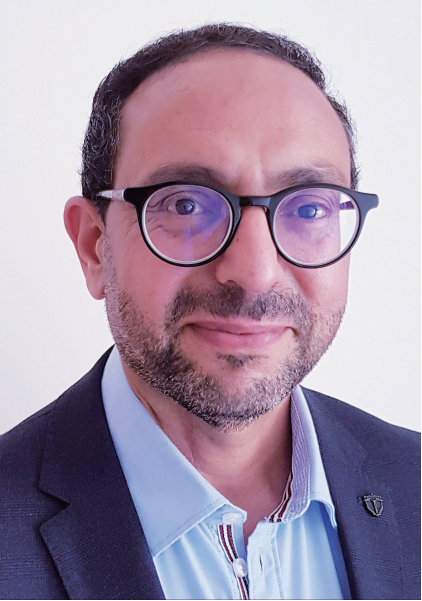}}]{Abderrahim Benslimane}
is Full Professor with the Laboratory of Computer Sciences at the Avignon University, France. He is Chair of the ComSoc Technical Committee of Communication and Information Security. He is EiC of Inderscience Int. J. of Multimedia Intelligence and Security (IJMIS), Area Editor of Security in IEEE IoT Journal, Area Editor of Wiley Security and Privacy Journal and Editorial Member of IEEE Wireless Communication Magazine, Elsevier Ad Hoc, IEEE Systems and Wireless Networks Journals. 
\end{IEEEbiography}\vspace{-5mm}

\begin{IEEEbiography}[{\includegraphics[width=1in,height=1.25in,clip,keepaspectratio]{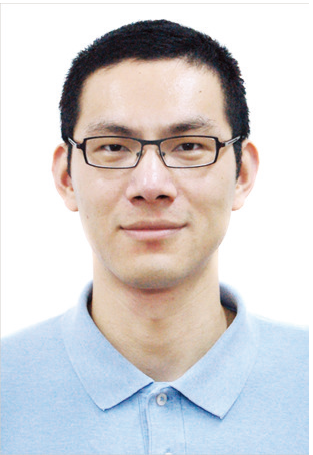}}]{Yuan Wu}
received the PhD degree in Electronic and Computer Engineering from the Hong Kong University of Science and Technology in 2010. He is currently an Associate Professor with the State Key Laboratory of Internet of Things for Smart City, University of Macau and also with the Department of Computer and Information Science, University of Macau. His research interests include resource management for wireless networks, green communications and computing, mobile edge computing and edge intelligence. He was a recipient of the Best Paper Award from the IEEE ICC2016, and the Best Paper Award from IEEE Technical Committee on Green Communications and Computing in 2017. He is currently on the Editorial Boards of {\scshape IEEE Transactions on Vehicular Technology}, {\scshape IEEE Transactions on Network Science and Engineering}, {\scshape IEEE Internet of Things Journal}, and {\scshape IEEE Open Journal of the Communications Society}.
\end{IEEEbiography}

\end{document}